\documentclass[a4paper,fleqn,usenatbib]{mnras}

\usepackage[T1]{fontenc}
\usepackage{ae,aecompl}

\usepackage{graphicx}	
\usepackage{amsmath}	
\usepackage{amssymb}	

\title[The Assembly of Virgo]{The Assembly of the Virgo Cluster, Traced by its Galaxy Halos }

\author[J. E. Taylor et al.]{
James E. Taylor,$^{1,2}$\thanks{E-mail: taylor@uwaterloo.ca}
Jihye Shin,$^{3}$
Nathalie N.-Q. Ouellette$^{4}$
and St\'{e}phane Courteau$^{5}$
\\
$^{1}$Waterloo Centre for Astrophysics, University of Waterloo, Waterloo, ON N2L 3G1 Canada\\
$^{2}$Department of Physics and Astronomy, University of Waterloo, 200 University Avenue West, Waterloo, Ontario N2L 3G1, Canada\\
$^{3}$Korea Astronomy and Space Science Institute (KASI), 776 Daedeokdae-ro, Yuseong-gu, Daejeon 34055, Korea\\ 
$^{4}$D\'{e}partement de physique, Universit\'{e} de Montr\'{e}al, Montr\'{e}al, QC H3C 3J7, Canada\\
$^{5}$Department for Physics, Engineering Physics and Astrophysics, Queen's University, Kingston, ON K7L 3N6, Canada\\
\vspace {-1cm}
}

\date{Accepted XXX. Received YYY; in original form ZZZ}

\pubyear{2019}

\begin{document}
\label{firstpage}
\pagerange{\pageref{firstpage}--\pageref{lastpage}}
\maketitle

\begin{abstract}
Kinematic studies have produced accurate measurements of the total dark matter mass and mean dark matter density
within the optical extent of galaxies, for large samples of objects. Here we consider theoretical predictions for the latter quantity, 
$\bar{\rho}_{\rm dm}$, measured within the isophotal radius $R_{23.5}$, for isolated halos with universal density profiles. 
Through a combination of empirical scaling relations, 
we show that $\bar{\rho}_{\rm dm}$ is expected to depend { weakly on halo mass and strongly on redshift}. When galaxy halos fall into larger groups or clusters they become tidally stripped, reducing 
their total dark matter mass, but this process is expected to preserve central density until an object is close to disruption. We confirm 
this with collisonless simulations of cluster formation, finding that subhalos have values of $\bar{\rho}_{\rm dm}$ close to the analytic predictions. This suggests that $\bar{\rho}_{\rm dm}$ may be a useful indicator of infall redshift onto the cluster.
We test this hypothesis with data from the {\it SHIVir} survey, which covers a reasonable fraction of the Virgo cluster. We find that
galaxies with high $\bar{\rho}_{\rm dm}$ do indeed trace the densest regions of the cluster, with a few notable exceptions. Samples selected by 
environment have higher densities at a significance of 3.5--4\,$\sigma$, while samples selected by density are more clustered at 3--3.5\,$\sigma$ significance.  
We conclude that halo density can be a powerful tracer of the assembly history of clusters and their member galaxies.
\end{abstract}




\section{Introduction}

In the standard picture of cosmological structure formation \cite[e.g.][]{White78}, galaxies form initially via gas cooling in individual dark matter haloes;
these haloes then merge hierarchically into groups and clusters of galaxies. In some cases, the merger of two haloes leads to a subsequent merger 
of their central galaxies; in other cases the smaller galaxy survives as a satellite occupying a self-bound `subhalo' that corresponds to a tidally truncated version 
of its original halo. In particular, all non-central cluster galaxies should occupy subhalos, whose inner structure is cut off from further growth or evolution. 
The dark matter distribution immediately surrounding these galaxies, out to distances of tens of kpc or more, should reflect the properties of the `field' halos 
they first formed in. 

Field halos themselves carry the imprint of a formation epoch at which they assembled some critical fraction of their mass in the density of the central region, expressed indirectly, for instance, via the { halo} concentration parameter \citep{NFW97}. 
{ Halo growth typically} progresses through rapid and slow phases \citep[e.g.][]{Wechsler02,Zhao09}. The central region forms and its density is set during the rapid growth phase, while subsequent slow growth leaves the central region unchanged. 

Combining these two ideas, that hierarchical merging preserves the central density of subhalos, and that this density reflects the epoch at which the subhalo formed, 
leads to the conclusion that an estimate of the dark matter densities in the subhalos of cluster member galaxies could reveal their history as independent objects before they fell into the cluster. Since the galaxy distribution within clusters 
is predicted to show radial gradients in infall epoch \citep[e.g.][]{Mamon04,Mahajan11,Oman13,Rhee17}, one should expect to see matching patterns in subhalo density.  

Gradients in stellar age or star formation rate have been measured in many different forms for galaxies in groups and clusters \citep[e.g.][]{Biviano97,Ellingson01,Kodama04,Balogh04a,Balogh04b,Rines05,Haines06,Haines09,Roediger11b,Hernandez14}. Stellar ages and star formation history need not track the 
assembly history of a galaxy, however \cite[e.g.][]{Hill17}. A dark matter halo may assemble early on, but not undergo much star formation until later; 
conversely, stars may form in small halos and merge together in a subsequent event that resets the density of the dark matter halo, 
but preserves the old stellar populations. The latter phenomenon, `dry merging', is suggested to be an essential part of the 
formation sequence for the most massive ellipticals that dominate clusters \cite[e.g.][]{Faber07,Khochfar09,Lidman12}. 
To demonstrate and quantify the age gradients expected from the theory of structure formation requires that they be measured through the properties of the dark matter component.
In particular, dark matter { subhalos} in the central regions of clusters must be shown to have higher densities than those at the periphery or in the field.

Probing the mass distribution around individual cluster galaxies is observationally challenging. 
Gravitational lensing provides a direct probe of the entire (projected) mass distribution, but becomes noisy
as the angular scale of interest decreases to the scale radius of typical galaxy halos (30--50 kpc,  corresponding to 5--10 
arcminutes at the distance of Virgo, the nearest large cluster) 
for relatively sparse ground-based shape catalogues. Furthermore, since lensing probes the projected mass
distribution, separating cluster and subhalo contributions can be challenging  \citep[e.g.][]{Gillis13,Li14,Sifon15,Li16,Niemiec17,Sifon18}.
The potential of lensing should increase dramatically with forthcoming space-based missions such as Euclid\footnote{http://sci.esa.int/euclid} 
and WFIRST\footnote{https://wfirst.gsfc.nasa.gov}. 

Kinematical mass determinations have the opposite problem; these tend to be most accurate on small scales (< 20 kpc) 
where they are dominated by the baryonic component. Kinematical masses determined using deep long-slit spectroscopy, 
however, may help bridge the gap between both scales and sample the intermediate range of 1 to 4 $R_{\rm e}$ where the baryon and dark-matter 
fractions are roughly equal. Using this method, the ``Spectroscopy and H-band Imaging of the Virgo cluster'' (SHIVir) survey \citep{Ouellette17} has 
measured kinematics out to large radii for almost 200 galaxies in the Virgo cluster. This dataset presents 
an opportunity to test the dark matter density in the subhalos of cluster members, over a large fraction of the cluster. 

First, however, we review theoretical expectations for the dark matter density at small radii around galaxies. 
These predictions combine the current understanding of halo properties (such as the infall redshift distribution and the concentration-mass relation), 
galaxy scaling relations (such as the size-stellar mass relation), and finally the stellar-to-halo-mass relation (SHMR). As a result, the expected pattern is
not obvious {\it a priori}, even in the absence of complicating effects such as adiabatic contraction.

The outline of the paper is as follows. In section \ref{sec:2}, we derive the expected scaling of halo density with stellar mass, using scaling relations, an analytic model, and collisionless cosmological simulations. In section \ref{sec:3}, we describe the SHIVir survey, and compare the densities measured by SHIVir to the predictions of the previous section. In section \ref{sec:4} we study { the variation of density} with environment across Virgo, showing that it does appear to trace the assembly history of the cluster. In section \ref{sec:5} we consider a few { complications and challenges to this interpretation, and their possible effects} on the result. Finally, in section \ref{sec:6} we conclude on the potential for dark matter density to reveal the evolutionary history of galaxies in other environments. Unless otherwise specified, we assume a Planck cosmology with  $\Omega_{{\rm m},0} = 0.315$, $\Omega_{\Lambda,0} = 0.685$, and $H_0 = 100 h$ km/s/Mpc with $h = 0.674$ \citep{Planck18}.

\section{Theoretical Predictions for Halo Density}
\label{sec:2}

\subsection{Expected { dependence on} stellar mass}
\label{subsec:2.1}

In the absence of baryonic effects, cosmological simulations indicate that dark matter should assemble into halos with a universal density profile, independent of mass, redshift or cosmology. This is most commonly approximated by the Navarro-Frenk-White (NFW -- \citealt{NFW96,NFW97}) profile
\begin{equation}
\rho(r) = \frac{\rho_0}{(r/r_{\rm s})(1 + r/r_{\rm s})^2} \, .
\label{eq:NFW}
\end{equation}
The NFW profile is characterized by a density $\rho_0$, a scale radius $r_{\rm s}$, and an outer `virial' radius $r_{\rm vir}$, 
or equivalently by { $\rho_{\rm 0}$}, $r_{\rm s}$, and a concentration parameter $c \equiv r_{\rm vir}/r_{\rm s}$. 
The outer, or virial, radius $r_{vir}$ is normally defined as the radius within which the mean density exceeds a reference background density 
$\rho_{\rm bg}$ by a specific (possibly redshift-dependent) factor $\Delta(z)$. Here we will take the background density $\rho_{\rm bg}$ 
to be equal to the critical density $\rho_{\rm c}(z) = 3H(z)^2/8\pi G$, and the overdensity to be the fixed value $\Delta = 200$.

More detailed studies at higher resolution \citep{Navarro04,Merritt06,Gao08} indicate that the mean halo density profile actually varies slightly with mass and redshift, and is better fit by the `Einasto' profile \citep{Einasto65}, which has an extra free shape parameter $\alpha$. The differences between the two profiles are largest on cluster-mass scales, however; for low-redshift galaxy halos where $\alpha \sim 0.15$, the NFW profile remains an accurate fit to within 10--20\%\  over the radial range 0.1--10 $r_{\rm s}$. Thus, for simplicity, we will assume an NFW profile in our predictions. 

Given the definitions above, the total mass of the { dark matter} halo is
\begin{equation}
M_{\rm h} = 4\pi \rho_0 r_{\rm s}^3 f(c)\, ,
\end{equation}
where 
\begin{equation}
f(x) = \ln(1+x) - \frac{x}{(1+x)}\, ,
\label{eq:fx}
\end{equation}
while the mean density of the halo is
\begin{equation}
\bar{\rho}(< r_{\rm vir}) = \frac{3M_{\rm h}}{4\pi r_{\rm vir}^3} = 3\rho_0\frac{f(c)}{c^3} = \Delta(z) \rho_{\rm bg}\, .
\end{equation}

Our comparisons with observations will focus on $R_{23.5}$, the radius at which the projected surface brightness profile drops below 23.5 mag arcsec$^{-2}$ { in the $i$-band}. Empirically, in the SHIVir data set discussed below, this radius scales with stellar mass roughly as $R_{23.5} \propto M_*^{1/3}$. \cite{Ouellette17} also showed that the 3D radius { $r_{\rm o} \sim R_{23.5} \propto M_*^{1/3}$}.\footnote{We note that the proportionality constant may be slightly different for early and late type galaxies -- see  \cite{Ouellette17} -- but we will ignore this complication here.} The density within this radius should then scale as {
\begin{eqnarray}
\bar{\rho}_{23.5} &=&  \frac{3M(<r_{\rm o})}{4 \pi r_{\rm o}^3}\\ 
&=& \left(  \frac{3M_{\rm h}}{4\pi r_{vir}^3} \right)\frac{f(x_{\rm o})}{f(c)}\frac{1}{(r_{\rm o}/r_{\rm vir})^3}\\
&=& \Delta (z) \rho_{\rm c}(z) \frac{f(x_{\rm o})}{f(c)}\left(\frac{r_{\rm vir}/r_{\rm s}}{r_{\rm o}/r_{\rm s}}\right)^3\\
&=& 200 \rho_{\rm c}(z) \frac{f(x_{\rm o})/x_{\rm o}^3}{f(c)/c^3}\, ,
\end{eqnarray}
where $x_{\rm o} = r_{\rm o}/r_{\rm s}$ and $f(x)$ given by equation \ref{eq:fx}.
}

For the range of halo masses relevant here, and  over the redshift range $z = 0$--3, {$x_{\rm o}$} is typically in the range $\sim 0.12$--0.25 (cf. Figs~\ref{fig:A1},\ref{fig:A2}), while the mean concentration parameter{, $c$, ranges  from 3 to} 20. In the range { for $x_{\rm o}$}, the function $f(x)/x^3$ scales as $x^{-\delta_1}$, with $\delta_1 = 1.1$--1.3 (cf. Fig.~\ref{fig:A3}), while in the range { for $c$},   $f(c)/c^3$ scales as $c^{-\delta_2}$, with $\delta_2 = 2.1$--{2.5} (cf. Fig.~\ref{fig:A4}). 
If we further assume that $M_{\rm h} \propto M_*^\beta$, where $\beta$ is the inverse logarithmic slope of the stellar-to-halo mass relation (SHMR), 
and that the concentration scales with mass as $c \sim M_{\rm h}^{-0.09}$ \cite[e.g.][]{Duffy08}, then since $r_{\rm o} \propto M_*^{1/3}$ and $r_{vir} \propto M_{\rm h}^{1/3}$, we have
\begin{eqnarray}
\bar{\rho}_{23.5} &\propto& \rho_{\rm c}(z)\, x^{-\delta_1} c^{\delta_2}\\  
&\propto& \rho_{\rm c}(z)\, r_{\rm o}^{-\delta_1} r_{\rm s}^{\delta_1} c^{\delta_2}\\  
&\propto& \rho_{\rm c}(z)\, r_{\rm o}^{-\delta_1} r_{\rm vir}^{\delta_1}\, c^{-\delta_1} c^{\delta_2}\\  
&\propto& \rho_{\rm c}(z)\, M_*^{-\delta_1/3} M_{\rm h}^{\delta_1/3} M_{\rm h}^{-0.09(\delta_1 - \delta_2)}\\  
&\propto& \rho_{\rm c}(z)\, M_*^{-\delta_1/3} M_*^{\beta\delta_1/3} M_*^{-0.09\beta(\delta_1 - \delta_2)}\\  
&\propto& \rho_{\rm c}(z)\, M_*^\gamma
\end{eqnarray}
where 
\begin{equation}
\gamma = \frac{d\ln\bar{\rho}}{d\ln M_*} = -\frac{\delta_1}{3}(1 - \beta) -0.09\beta(\delta_1 - \delta_2)\,.
\end{equation}

Using the analytic SHMR of \cite{Behroozi13}, we find that $\beta$ ranges from a value of $\sim 0.5$ at low masses ($\log_{10} M_* = $\,8.5--10), to  $\sim2$ or more at high masses ($\log_{10} M_* \sim$\,11). Combining this with the values $\delta_1 \sim 1.2$ and $\delta_2 \sim 2.3$, we predict that $\gamma$ should vary from $-0.15$
at low mass to $+0.6$ at high mass. The fact that this slope { ranges} from mildly negative to slightly positive over the broad range of stellar mass corresponding to typical 
galaxies indicates that {\it the overall trend in mean dark matter density with stellar mass should be approximately flat} at a given redshift. 

The main redshift dependence of $\bar{\rho}_{23.5}$ is through the critical density, which scales as $H(z)^2$, varying by a factor of 20 from $z=0$ to $z=3$ in our chosen cosmology. The actual range of density may be reduced by the redshift dependence of the concentration-mass relation, however; the mean concentration drops by a factor of 2--3, depending on the halo mass, between $z=0$ to $z=3$. Since $\bar{\rho}_{23.5} \propto c^{\delta_2} \sim c^{2.3}$, we expect the density range to be reduced by a factor of (2--3)$^{2.3} \sim 4$--8, so the range may only vary by 2--4 from $z=0$ to $z=3$. We will show that this is indeed the range predicted by our more detailed analytic model (cf.~Fig.\ref{fig:2} below).

This scaling argument ignores many complications, however, including the scatter in the $r_{\rm o}$--$M_*$ and $M_*$--$M_{\rm h}$ relations and the curvature in the latter. To account for these, we will proceed to construct a more realistic, analytic model for the variation of $\rho_{23.5}$ with $M_*$.

\subsection{A more detailed analytic model}
\label{subsec:2.2}
 
We now consider a more detailed model for the { dependence} of mean dark { matter} density on stellar mass. The model includes three main components:
\begin{enumerate}
\item A stellar-to-halo mass relationship (SHMR), or its inverse, the halo-to-stellar mass relationship (HSMR), to convert between observed stellar masses and corresponding dark matter halo masses, and a description of the scatter in this relationship.
\item A mean concentration-mass-redshift relation, and a description of the scatter in this relationship.
\item An empirical relationship between $R_{23.5}$ or $r_{\rm o}$ and stellar mass, and a description of the scatter in this relationship.
\end{enumerate}

For the HSMR, we use the functional form proposed by  \cite{Behroozi10} 
\begin{displaymath}
\log_{10}(M_h) = \hspace{0.65\columnwidth}
 \end{displaymath}
 \vspace{-3ex}
$$\quad \log_{10}(M_1) + \beta\,\log_{10}\left(\frac{M_\ast}{M_{\ast,0}}\right) +
 \frac{\left(\frac{M_\ast}{M_{\ast,0}}\right)^\delta}{1 + \left(\frac{M_{\ast}}{M_{\ast,0}}\right)^{-\gamma}} - \frac{1}{2},\hspace{0.05\columnwidth}$$
We use the parameter values [$\log(M_1/M_\odot)$, $\log(M_{\ast,0}/M_\odot)$, $\beta$, $\delta$, and $\gamma$] = [12.45, 10.35, 0.39, 0.4, 1.0], determined in \cite{Grossauer15} by fitting the stellar mass function of the Next Generation Virgo Cluster Survey \citep{NGVS}, over the range $\log(M_*/M_\odot)$ = 5.0--10.5.
They note that $\delta$ and $\gamma$ are not well constrained by these data; after some experimentation we have adopted the values [$\delta$, $\gamma$] = [0.2, 1.0] instead, as these produce a more reasonable range of halo masses at large stellar mass, given the bias introduced by scatter in the relation.

SHMR models normally assume log-normal scatter in stellar mass at fixed halo mass, $\sigma_{M_*}$. Fitting to data at low redshift, \cite{Leauthaud12} estimated that $\sigma_{M_*}  \sim 0.2$. Here we will include this scatter approximately by adding log-normal scatter $\sigma_{M_{\rm h}} = 0.2$ to the HSMR. This approach is not technically correct as it biases the mean relation, as discussed in \cite{Leauthaud12}, but the bias can be accounted for by adjusting the value of $\delta$ and $\gamma$ as discussed above. 

For the concentration-mass redshift relation, we use the model of \cite{Klypin16}
\begin{equation}
c(M_{\rm h}) =  c_0\,\left(\frac{M_{\rm h}}{M_{12}}\right)^{-\gamma}\left[1 + \left(\frac{M_{\rm h}}{M_0}\right)^{0.4}\right]\, , 
\end{equation}
where $M_{12} = 10^{12} h^{-1} M_\odot$. The authors list values of the constants $c_0$, $\gamma$ and $M_0$ for the Planck cosmology at various discrete redshifts in their Table 2. We use the following approximations to interpolate these to other redshifts
\begin{eqnarray}
 c_0(z) &=& 0.5 + 7.5/(1 +z)\, ;\\
\gamma(z) &=& 0.078 + 0.05\exp(-z)\,;\\
\log_{10} M_0(z) &=& -1.5 + 8.0/(1+z)\,.
\end{eqnarray}
The scatter in the concentration-mass relation is taken to be log-normal, with a fixed amplitude $\sigma_{\log c} = 0.16$, based on the results of \cite{Diemer15}.

Finally, we determine the relationship between $R_{\rm 23.5}$ and stellar mass empirically from the SHIVir data (discussed in Section \ref{sec:3} below). Fig. \ref{fig:1} shows $R_{\rm 23.5}$ versus $M_*$ for the SHIVir sample (green squares). The solid line shows the single power-law approximation, $R_{\rm 23.5} \sim r_{\rm o} \propto M_*^{1/3}$, used to derive the scaling relation in the previous section. While this is a reasonable fit to the data, the slope appears to show some curvature, flattening at low stellar mass and increasing at high mass (see also \citealt{Ouellette17}). We adopt the following model for the mean relation
\begin{equation}
R_{\rm 23.5}\sim r_o = \left[1.0 + 5.5\left(\frac{M_*}{10^{10} M_\odot}\right)^{0.45}\right]\, {\rm kpc}\,,
\end{equation}
while we take the scatter to be log-normal with a mass-dependent  r.m.s.~value 
\begin{equation}
\sigma(R_{23.5}) = 0.2\left[\log_{10} (M_*/M_\odot) - 7.0\right]\,.
\end{equation}
The blue background points in Fig.~\ref{fig:1} show a random sample of values with this mean and scatter. Overall, they provide a fairly good description of the data, 
although there are somewhat more outliers than predicted by the model.

\begin{figure}
	\includegraphics[width=\columnwidth]{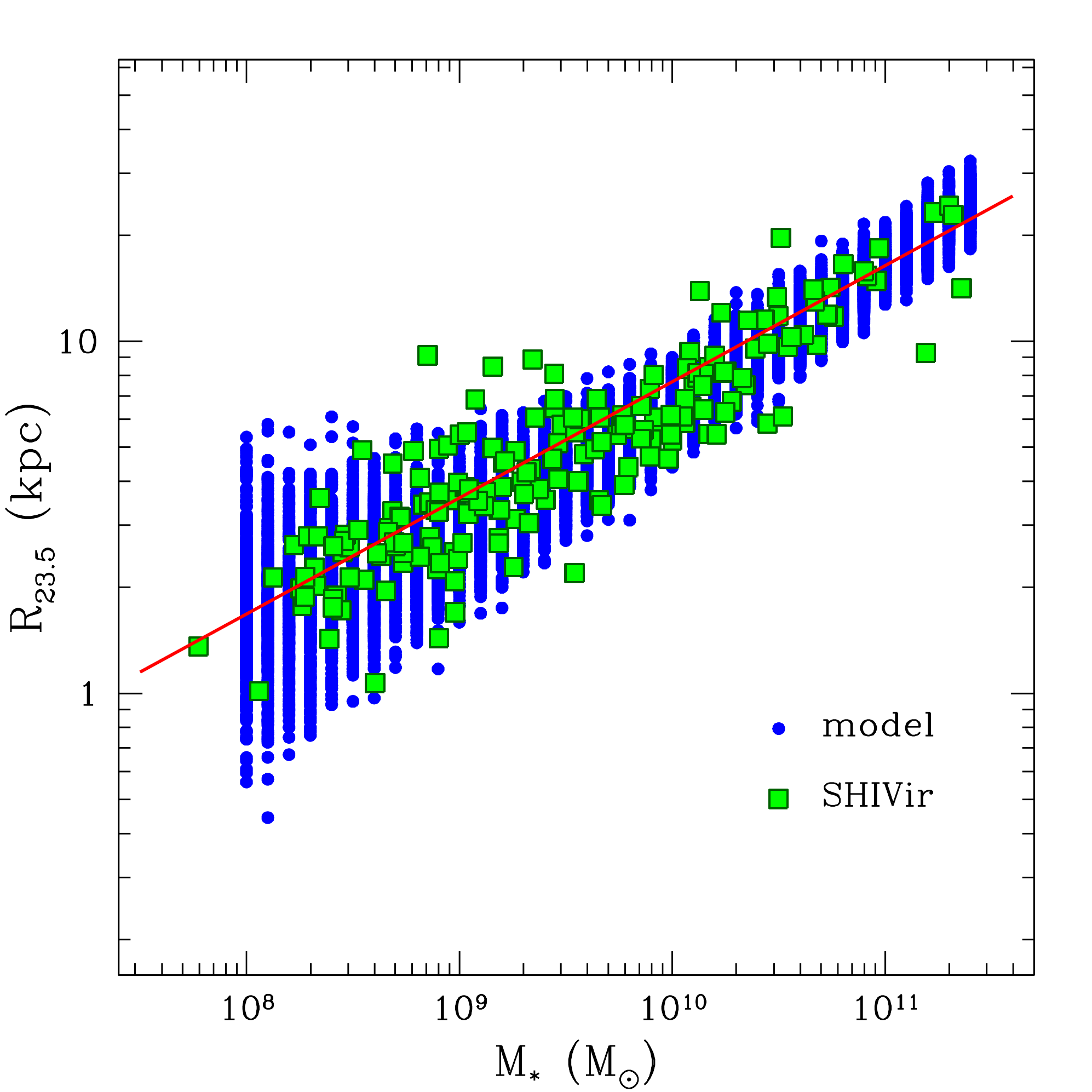}
    \caption{Outer isophotal radius versus stellar mass, for the SHIVir sample (green squares). The line indicates a power law of slope $^1/_3$, while the blue circles are a random sample with a mean and r.m.s. given by the model described in the text.}
    \label{fig:1}
\end{figure}

\begin{figure}
	\vspace{-2cm}
	\includegraphics[width=\columnwidth]{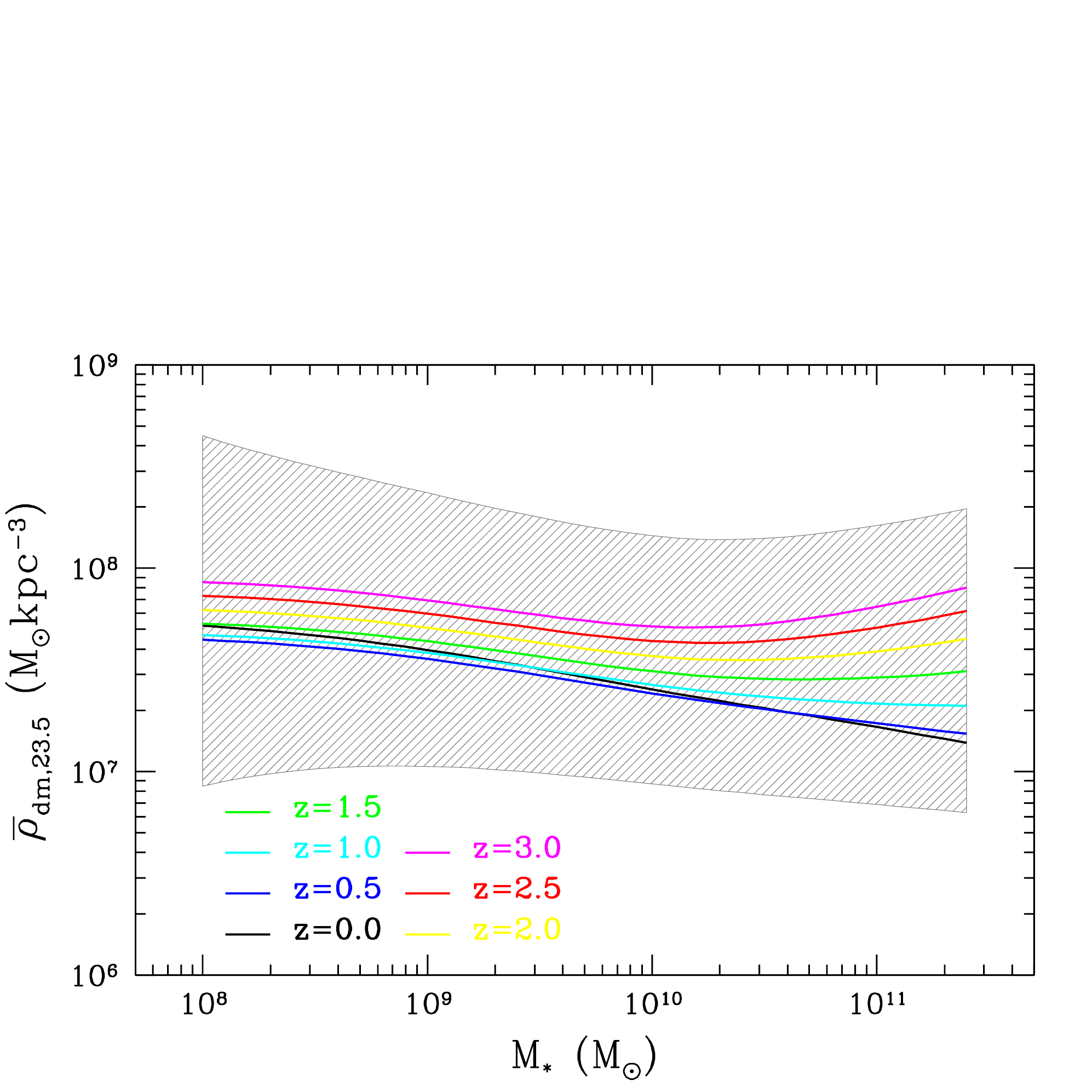}
    \caption{Mean dark matter density within the 3D radius $r_{\rm o}$ corresponding to the isophotal radius $R_{23.5}$, $\bar{\rho}_{23.5}$, versus stellar mass. The solid curves show the mean relations for $z=0$ to $z=3$, as labelled, while the hatched region shows the 2-$\sigma$ range expected, given the models for scatter discussed in the text.
    }
    \label{fig:2}
\end{figure}

Combining these three elements, given a stellar mass and redshift, we can predict a halo mass and virial radius, a concentration and scale radius, and finally a radius $r_{\rm o}$. From these it is straightforward to calculate the mass of dark matter enclosed within $r_{\rm o}$, and thus the mean density $\bar{\rho}_{23.5}$. Fig.~\ref{fig:2} shows the resulting mean relations (solid curves, for $z=0$ to $z=3$ from bottom to top) and the 2-$\sigma$ range expected given the models for scatter given above. The total scatter has a r.m.s. of $\sigma_{\log\rho} \sim 0.24$, and is dominated by the scatter in concentration, and then in $R_{23.5}$. The contribution due to scatter in the SHMR is not significant, though it may be underestimated slightly since we apply it to the HSMR rather than the SHMR.

As expected, the trend of mean density with stellar mass is generally fairly flat, changing from very slightly negative at low mass to positive at high mass, depending on redshift. While the mean relation increases by a factor of 2--5, depending on the mass range, from $z=0$ to $z=3$, the range of the 2-$\sigma$ scatter corresponds to a factor of 30--40.  Overall, the analytic model validates the estimate derived in the previous section.

Both the scaling relation derived above and the more detailed analytic model indicate that mean density should be relatively independent of stellar mass, 
and should increase systematically with redshift. On the other hand, it seems that the halo-to-halo scatter may obscure the trend with redshift, since the former is a factor of 30--40,
whereas the latter is only a factor or 2--5. We should reconsider the meaning of ``redshift" in this context, however. Our analytic model assumes the virial density appropriate for a given redshift in order to calculate the virial radius, and uses the mean (field) concentration-mass relation at that redshift to calculate the scale radius and density normalization $\rho_0$. Thus, it assumes that a galaxy occupies a typical field halo at that redshift. If galaxy halos do indeed have typical field properties when they first fall in to the cluster, and tidal stripping preserves their central density profile, then our predictions should be correct if we take the redshift in the analytic model to be the infall redshift of the galaxy. 

There are several complications to this. First, the halos of cluster galaxies formed and evolved in an unusually dense environment even before they fell in to the cluster, and thus will have assembled their mass earlier than average. This in turn implies that they will have higher-than-average concentrations, since concentration is related to formation time \citep[e.g.][]{NFW97, Wechsler02, Zhao09}. Thus, mean density may vary more strongly with redshift than predicted by the analytic model. 
Second, the assumption that cluster galaxies preserve their infall properties, or at least the value of $\bar{\rho}_{23.5}$ calculated from these, may be  unrealistic. Tidal stripping could reduce the inner density of subhalos slightly, although the magnitude of the effect is not expected to be large (e.g. the density should drop by only 20\%\ for 50\%\ mass loss -- \citealt{Hayashi03}; see also \citealt{Drakos17}). Major mergers between subhalos may have more significant effects, however, and could be common for more massive galaxies as they are accreted through multiple stages of hierarchical merging in groups, before falling in to the final cluster. 

{ Finally, galaxies may modify the dark matter potential of their halo or subhalo via baryonic effects such as adiabatic contraction; these effects are discussed in} Section \ref{sec:5}.

\subsection{Results from Cosmological Simulations} 
\label{subsec:2.3}

To test how the inner densities of subhalos compare to those of field halos, we have examined substructure in a set of high-resolution cosmological dark-matter-only simulations of cluster formation. Subhalo orbits and merger rates in these simulations will be discussed further in forthcoming work (Shin et al.~in preparation). 

The original simulations were carried out earlier, in a slight variant of our Planck cosmology, with parameters  $\Omega_{\rm m,0} = 0.3$, $\Omega_{\Lambda,0} = 0.7$, $H_0 = 68$ km/s/Mpc, $\sigma_8 = 0.82$ and $n = 0.96$  (with the exception of $\Omega_{\rm m,0} = 0.3$, these are within about 2\,$\sigma$ of the Planck 2018 values). Multi-scale initial conditions were generated using the package MUSIC \citep{MUSIC}, with position and velocity displacements calculated using second-order Lagrangian perturbation theory, and a transfer function calculated using the CAMB package \citep{CAMB}\footnote{Available at http://camb.info}. The largest box, 140\,$h^{-1}$\,Mpc on a side, was evolved from $z=200$ down to $z=0$, using the $N$-body code GADGET2 \citep{GADGET2}, with particles of mass $1.7\times 10^9 M_\odot/h$ and a softening length of 5.469 kpc/$h$. 

Halos were identified in the final output at $z=0$, using a parallel {friends}-of-friends (FoF) algorithm \citep{Kim06}. A subset of these were chosen for resimulation at higher resolution, using several levels of refinement, and following the prescription of \cite{Onorbe14} to determine the minimum (ellipsoidal) Lagrangian volume to resimulate. The highest resolution resimulations had a particle mass $m_{\rm p} = 3.32\times 10^6 M_\odot/h$ and a softening length of ${ \epsilon} =$ 0.683 kpc/$h$. From this final set of resimulations, we saved 120 outputs between redshifts $z=9$ and $z=0$. The Amiga Halo Finder (AHF) \citep{AMIGA} was used to identify halos and subhalos in each output, and generate merger trees for all subhalos within the virial radius at the final output. 

In this work we considered in detail the substructure within three { high-resolution} clusters, { each with $m_{\rm p} = 3.32\times 10^6 M_\odot/h$ and $\epsilon =$ 0.683 kpc/$h$}. These had final virial masses (defined in terms of the spherical collapse overdensity) 
in the range 1.2--1.5$\times 10^{14\,}M_\odot$, whereas the estimated mass of Virgo is {roughly 3} times this { (4.2$\times10^{14} M_\odot$ --} \citealt{McLaughlin99}; see \citealt{Grossauer15} for a brief discussion of the uncertainties on this estimate). To correct for this, we rescaled the units of the simulations, scaling masses by {a factor of 3, and radii by a factor of $3^{1/3}$}, such that the mean density of each halo and subhalo remains the same.
{ For reasons of computational efficiency, the cluster sample chosen for resimulation was selected to have small Lagrangian volumes at a given mass in the original simulation, and may not be fully representative. While we did not see any superficial trends between final cluster properties and initial Lagrangian volume,} all three clusters considered here have fairly recent formation times, with significant accretion after z = 0.2 (especially one, which experiences a major merger around $z\sim0.1$). In future work, we will explore the effect of formation history on subhalo density for a larger { and more diverse} sample of clusters.

The three clusters each had more than 5000 well-resolved subhalos within their virial radius. For each subhalo, we measured the mass enclosed within a set of shells of fixed physical radius ranging from 1 to 32 kpc (before scaling), centred around the subhalo centre defined by AHF. We sorted subhalos by the mass enclosed within 16 kpc,
eliminating in the process any subhalos that had been tidally stripped to this radius or less. We then selected the 190 subhalos with the largest masses within 16 kpc, $M_{16}$, to match the 190 objects in the SHIVir sample (discussed in section \ref{sec:3} below). Stellar masses and outer galaxy radii $r_{\rm o}$ were assigned to these objects by abundance matching, that is assuming a monotonic relationship between stellar mass and the subhalo mass $M_{16}$. Finally, interpolating between the masses within fixed radii, we determined the mass enclosed within the outer radius $r_{\rm o}$ assigned to each object, and thus the mean dark matter density equivalent to $\bar{\rho}_{23.5}$.

As discussed further in Section \ref{sec:3}, { the 190 SHIVir objects sample a representative fraction of the most massive galaxies in Virgo}. Furthermore, given the scatter in the SHMR, the most massive galaxies will not necessarily occupy the most massive subhalos. We experimented with abundance matching on a fraction of a larger sample of subhalos, and found this had only a minimal effect on our results, reducing the mean densities very slightly, as enclosed masses were being measured in less massive objects. We also experimented with sorting subhalos by $M_{32}$, the mass enclosed within 32 kpc, and found this had no significant effect on the results { either}.

\begin{figure}
	\vspace{-3.5cm}
	\includegraphics[width=\columnwidth]{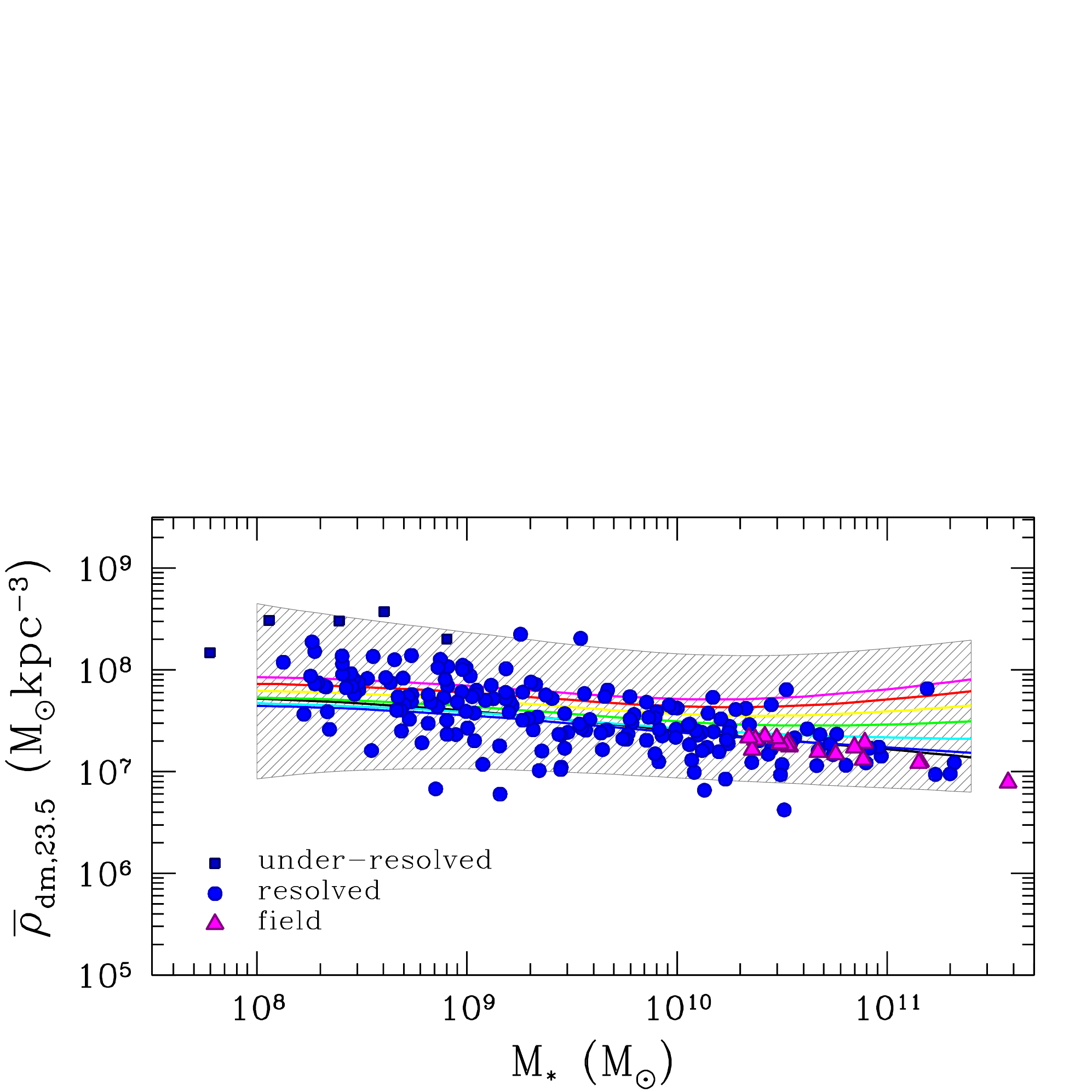}
    \caption{Mean dark matter density within $r_{\rm o}$ versus stellar mass, for the analytic model and the simulations. The solid curves and hatched region show the predictions of the analytic model, as in Fig.~\ref{fig:2}. The (blue) circles show values calculated for simulated subhalos, using the method described in the text. {Dark blue squares show under-resolved subhalos where $r_{\rm o}$ is less than the softening length. Magenta triangles show values for field halos in the high-resolution region of the simulation.}
    }
     \label{fig:3}
\end{figure}

Fig.~\ref{fig:3} shows the comparison between the analytic model for field halos discussed previously, with the 2-$\sigma$ range of scatter expected, and the results for simulated subhalos. { The (blue) circles show values calculated for simulated subhalos within one cluster, with units scaled as described above to correct for the lower mass of the cluster relative to Virgo. Dark blue squares show subhalos that are not properly resolved in the simulation, in the sense that $r_{\rm o}$ is less than the softening length. The high-resolution volume of the simulation also contains a few field halos outside the cluster; these are shown by the magenta triangles. (Note that these objects did not have their masses and radii rescaled, since they are not subhalos within a larger system. Stellar mass was assigned based on their halo mass, using the SHMR discussed in section \ref{subsec:2.2}.)}

Overall, there is {good} agreement between the simulation results and the analytic predictions; subhalo density within $r_{\rm o}$ depends only weakly on mass, and has roughly the normalization and scatter predicted by the analytic models. {There is a visible trend in the subhalo distribution of decreasing density with increasing mass, which is is expected as the more massive subhalos have formed and merged more recently, and should therefore match the low-redshift mean field relations. There are also a few outliers in the subhalo sample beyond the predicted 2-$\sigma$ contour, both at high and at low density. These are systems with particularly large or small values of $r_{\rm o}$, that is they correspond directly to the outliers in Fig.~\ref{fig:1}. These outliers reappear in Fig.~\ref{fig:3} because we have used the {\it observed} values of $r_{\rm o}$ from the SHIVir sample to set where to measure $\bar{\rho}_{23.5}$ in the simulations, abundance matching between individual SHIVir galaxies and the subhalos, as explained above.

Given our adopted scaling, at the largest stellar masses ($M_* > 4\times 10^{10} M_\odot$), the mean relation between density and stellar mass may be 20--30\%\ lower in the simulations than in the analytic model. Numerical effects such as relaxation \citep{Diemand04} could explain some of this offset, although we would expect these effects to dominate at smaller masses. (The less massive subhalos are about as dense as expected, though since they generally formed earlier, they should be compared with the higher-redshift predictions.) Our analytic model also ignores the effect of tidal stripping on the inner density. Stripping could affect the central density in principle, although detailed work by \cite{Hayashi03} modelling the effect of stripping on the density profile shows that, given the fractional mass loss rates expected for typical subhalos, the reduction in central density is only a factor of 10--20\%.

The few field halos present in the high-resolution volume of the simulation (magenta triangles) provide additional evidence for this, since they should not suffer from tidal stripping. In general, they have densities consistent with the predictions for isolated systems at $z=0$. There are, however, too few of them to constrain the mean density precisely.

Overall, we conclude that taken together, the analytic and numerical results provide a reasonable range of predictions with which to compare the observational results from SHIVir.}

\section{Halo Densities in the SHIVir Survey}
\label{sec:3}

Having established that the inner density of dark matter halos or subhalos, measured at the outer radius of the galaxy $r_{\rm o}$, should depend only weakly on halo or stellar mass, but vary systematically with infall or formation redshift, we now consider observational determinations of this quantity in Virgo, using data from the SHIVir survey.

\subsection{The SHIVir Survey} 
\label{subsec:3.1}

The Spectroscopy and H-band Imaging of the Virgo cluster (SHIVir) survey is an ongoing photometric and spectroscopic survey of the Virgo cluster, { aimed at understanding} galaxy dynamics and scaling relations over a broad range of stellar mass, age and metallicity, throughout the cluster. The data include optical and near-infrared photometry in multiple bands, as well as long-slit spectroscopy. An overview of the survey, as well as the data, analysis methods, and the results used here are presented in detail in \cite[][Ouellette et al.~in prep.]{Ouellette17}.

The survey sample draws from the set of the galaxies in the magnitude-limited Virgo Cluster Catalog (\citealt{Binggeli85}, hereafter VCC) for which optical imaging was available in the Sloan Digital Sky Survey 6th Data Release \citep{SDSS6}. 
Of these, a morphologically representative subsample of 286 galaxies were imaged at H-band \citep{McDonald09}. Deep long-slit spectroscopy was acquired for 140 of these objects, including 37 in the ACSVCS \citep{Cote04}, and 103 in dedicated SHIVir programmes described in \cite{Ouellette17}. These were combined with reliable linewidths from the literature for 50 other objects, to give a final sample of 190 galaxies. The sample covers more than half the spatial extent of Virgo, out to and even beyond its estimated virial radius. In particular, the areas around M87 and M60 are well sampled, although the area around M49 is incomplete, as discussed in Section \ref{sec:4}.

Dynamics for the SHIVir sample were determined from red emission features (H$\alpha$, [N$_{\mathrm{II}}$]) for gas-rich systems, and blue absorption features for gas-poor systems. Extended rotation curves, corrected for inclination, were determined for 46 objects. Using these, \cite{Ouellette17} established that a velocity measured at $R_{23.5}$, the projected radius corresponding to the $i$-band 23.5 mag arcsec$^{-2}$ isophotal level, gives the smallest scatter in the stellar-to-halo mass relation. For this reason, all structural quantities for galaxies were measured within R$_{23.5}$. Similarly, velocity dispersions were calculated for 135 early-type galaxies at various multiples of the effective radius $R_{\rm e}$, and the value at one effective radius was used to estimate the dynamical mass within $R_{23.5}$. 
(While the ratio $R_{23.5}/R_{\rm e}$ for SHIVir galaxies is $\sim$2 on average for  SHIVir galaxies, and varies considerably from system to system, the change in the velocity dispersion between the two radii is negligible, so the value measured at 1 $R_{\rm e}$ was used.) The stellar mass within the same radius was also estimated for both samples, using the integrated luminosity and a stellar mass-to-light ratio based on the multi-band optical and near infrared colours \citep{Roediger11a,Roediger11b}. Gas masses, where available, were taken from the ALFALFA $\alpha$.40 catalogue \citep{Haynes11}. Subtracting stellar mass and gas mass from the dynamical mass, the total dark matter mass within $R_{23.5}$ was estimated for all 190 objects.

It is not clear exactly what uncertainties to associate with these measurements. Typical systematic errors in the $R_{23.5}$ measurements are in the range of 15-20\%, which is negligible relative to the scatter in the $R_{23.5}$--$M_*$ relation (shown previously in Fig.~\ref{fig:1}). The error in the enclosed dark matter mass is assumed to be dominated by the error in the stellar mass that is subtracted from the total dynamical mass; we take this to be a fixed 0.3 dex (or a factor of 2), although in practice it will depend on magnitude and galaxy type. Thus we will take the error in the mean dark matter density to be $\sigma_{\bar{\rho}} = 3M_{*}/(2\pi R_{23.5}^3)$.
This uncertainty is mainly a concern for objects with low dynamical and/or dark matter masses, where the dark contribution to the potential is only marginally detected. 

There are several other error sources that may affect our estimates. Not all late-type galaxies in the sample had reliable distance estimates; in cases where these were lacking, a distance of 16.5 Mpc was assumed. Systematic uncertainties in the stellar mass due to the choice of initial mass function can exceed 0.3 dex, while  other modelling choices may contribute an additional 0.2 dex  to the error budget \citep{Roediger15}. Since larger errors in stellar or dark matter mass would reduce the significance of any radial trend in dark matter density with environment, correcting for these errors would actually increase the significance of the result reported below. Thus we make the conservative assumption that the errors are only 0.3 dex in stellar mass. 

\subsection{Comparison to Theory} 
\label{subsec:3.2}

\begin{figure*}
	\vspace{-7cm}
	\includegraphics[width=1.99\columnwidth]{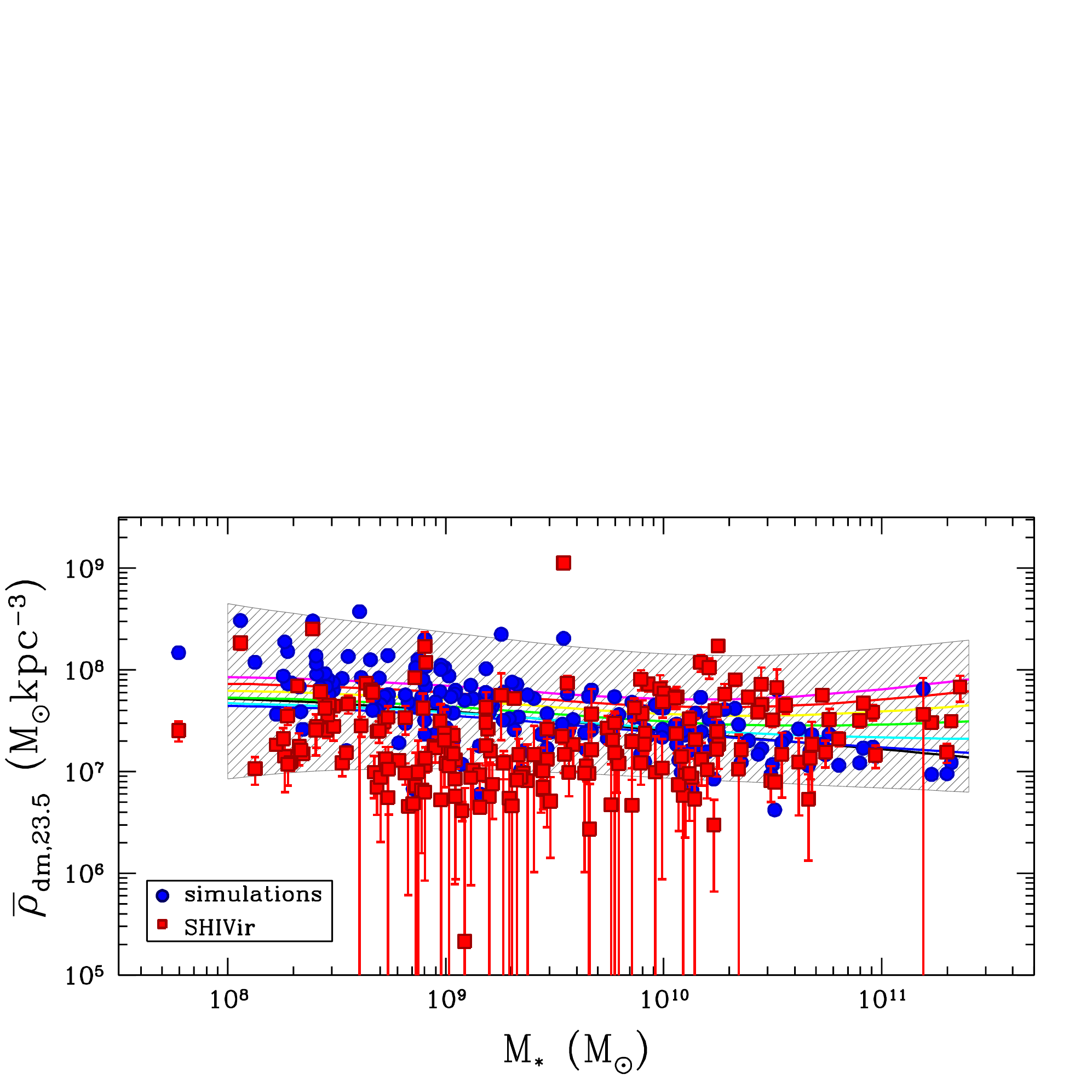}
    \caption{Mean dark matter density within $r_{\rm o}$ versus stellar mass, for the analytic model, the simulations, and the SHIVir sample. The solid curves and hatched region show the predictions of the analytic model and expected 2-$\sigma$ scatter, as in Fig.~\ref{fig:2}. The (blue) circles show values calculated for simulated subhalos, as in Fig.~\ref{fig:3}. The (red) squares with vertical error bars indicate the values measured for the SHIVir sample. Errors in $M_{*}$ are not shown for clarity, but are assumed to be approximately 0.3 dex at all stellar masses.}
    \label{fig:4}
\end{figure*}

Figure \ref{fig:4} shows the mean dark matter density estimated for the 190 SHIVir galaxies (red squares with error bars), compared to the theoretical predictions of the analytic model (solid curves and shaded region, as in Fig.~\ref{fig:2}) and the cluster simulations (blue points, as in Fig.~\ref{fig:3}). Overall, the agreement between the observational results and the theoretical predictions is reasonable. The mean dark matter density is similar ($\sim $2--3 $\times 10^{7} M_\odot/$kpc$^3$), as is the scatter (a factor of 5--10), and there is no particular trend {with} stellar mass. The observations are particularly close to the subhalo distribution for stellar masses $< 2\times 10^{10} M_\odot$; above this mass, the observed systems are 3--4 times denser, although their densities match those predicted by the analytic model for $z = 2$--3. At intermediate to low masses ($M_* \sim 10^{9}M_\odot$), the observed systems appear to be slightly less dense (by a factor of $\sim2$) than the simulated subhalos, although the large observational errors at low density complicate the comparison. Finally, there is one outlier with a very large density, the compact elliptical NGC4486B (VCC1297), which is four times denser than the next densest system. The velocity and size measurements for this object appear to be correct, so the reason for its unusual density remains unclear.

Morphological classifications are also available for the SHIVir sample, from the GOLDMine database \citep{Gavazzi03}. As explained in the GOLDMine glossary\footnote{http://goldmine.mib.infn.it/glossary.html}, the numerical Hubble T-types given in the database are a simplified version of the classification scheme used in the Virgo Cluster Catalog \citep{Binggeli85}. Classes $-3$ to $-1$ cover the range from dS0 to dE, class 0 is E or E/S0, class 1 is S0, classes 2--9 cover the range S0a/Sa--Sd, classes 10--13 cover Sdm--Im--Pec, 14--17 are blue compact dwarfs, while 18--20 are other dwarf types or unknown. In the SHIVir sample, early-type galaxies have a Hubble type between $-3$ and 2 inclusively, while late-type galaxies range between 3 and 20 inclusively.

\begin{figure}
	\includegraphics[width=0.99\columnwidth]{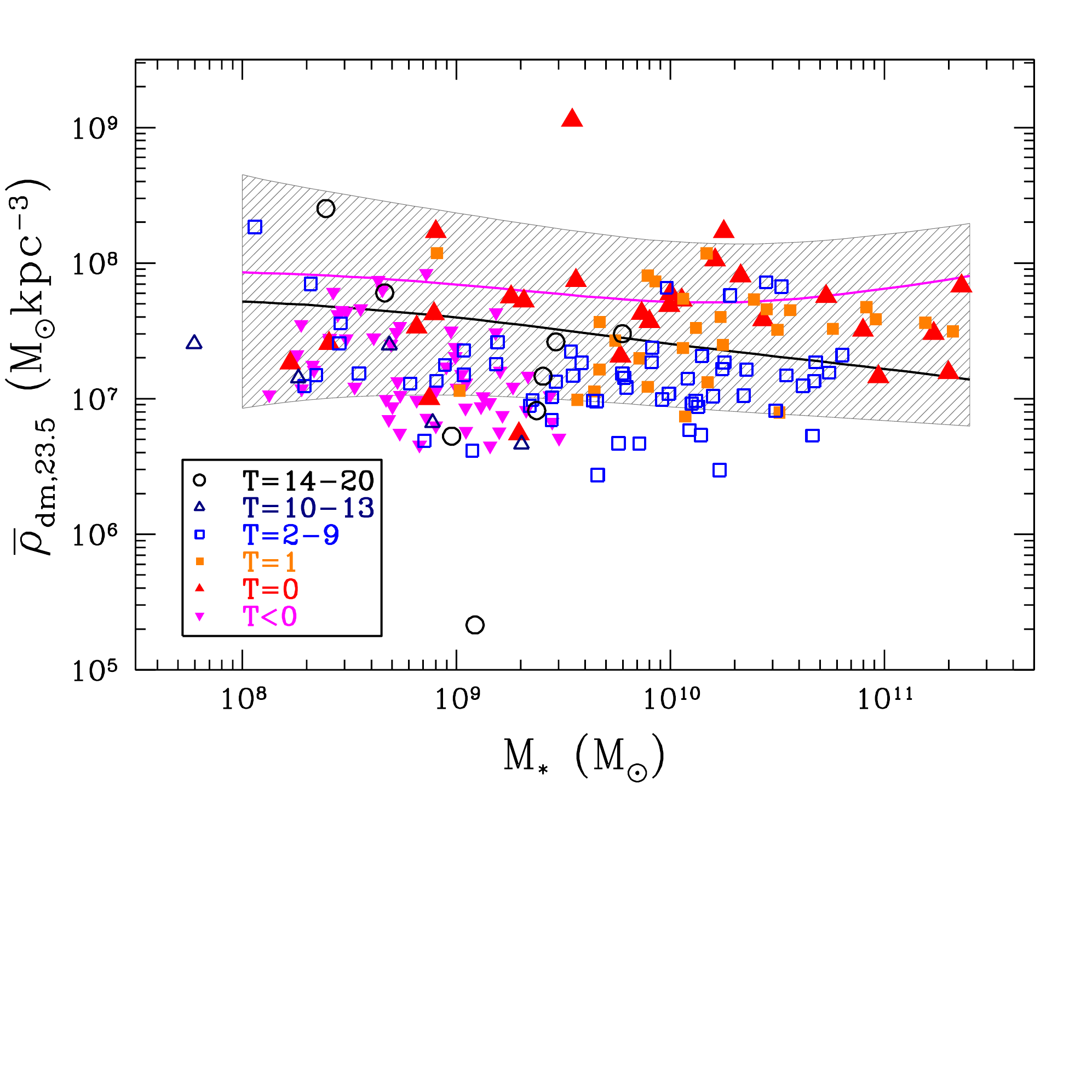}
    \vspace{-2.5cm}
    \caption{Mean dark matter density within $r_{\rm o}$ versus stellar mass for the SHIVir sample, as in Fig. \ref{fig:4}, but split by morphological T-type. Symbols indicate T-type $-3$ to $-1$ (inverted pink triangles), 0 (upright red triangles), 1 (orange squares), { 2--9 (open blue squares), 10--13 (open blue triangles), and 14--20 (open black circles).}}
    \label{fig:5}
\end{figure}

Fig.~\ref{fig:5} shows the trends in density versus stellar mass, splitting the sample by morphological type. Dwarf ellipticals (T-type $-3$ to $-1$, inverted pink triangles) generally have slightly lower densities than expected, particularly at higher stellar masses. Giant ellipticals and S0s (T-types 0--1, upright red triangles and orange squares respectively) have densities above the mean value predicted by the analytic model, especially at higher stellar masses. Spirals (T-types 2--{9}, open blue squares) lie below the mean relation, except for a few exceptions at large or small stellar masses. Finally irregulars, blue compact galaxies and the other dwarf classes generally have densities below the mean relation. Overall, although the different morphological classes tend to have systematically higher or lower dark matter densities, there are enough exceptions in every class to cause us to seek a different explanation for the observed range in density.

Comparing to the model predictions, the range in dark matter density measured in the SHIVir sample suggests that we may be seeing systematic differences in formation epoch. Since the location of objects in the cluster should be correlated with their infall and formation redshifts, it is interesting to consider dark matter density with respect to spatial location within the cluster, to see if any systematic patterns are visible. We will {address} this question next.

\section{Patterns in the Spatial Distribution within Virgo}
\label{sec:4}

\subsection{Known Virgo Substructure} 
\label{subsec:4.1}

\begin{figure}
	\includegraphics[width=\columnwidth]{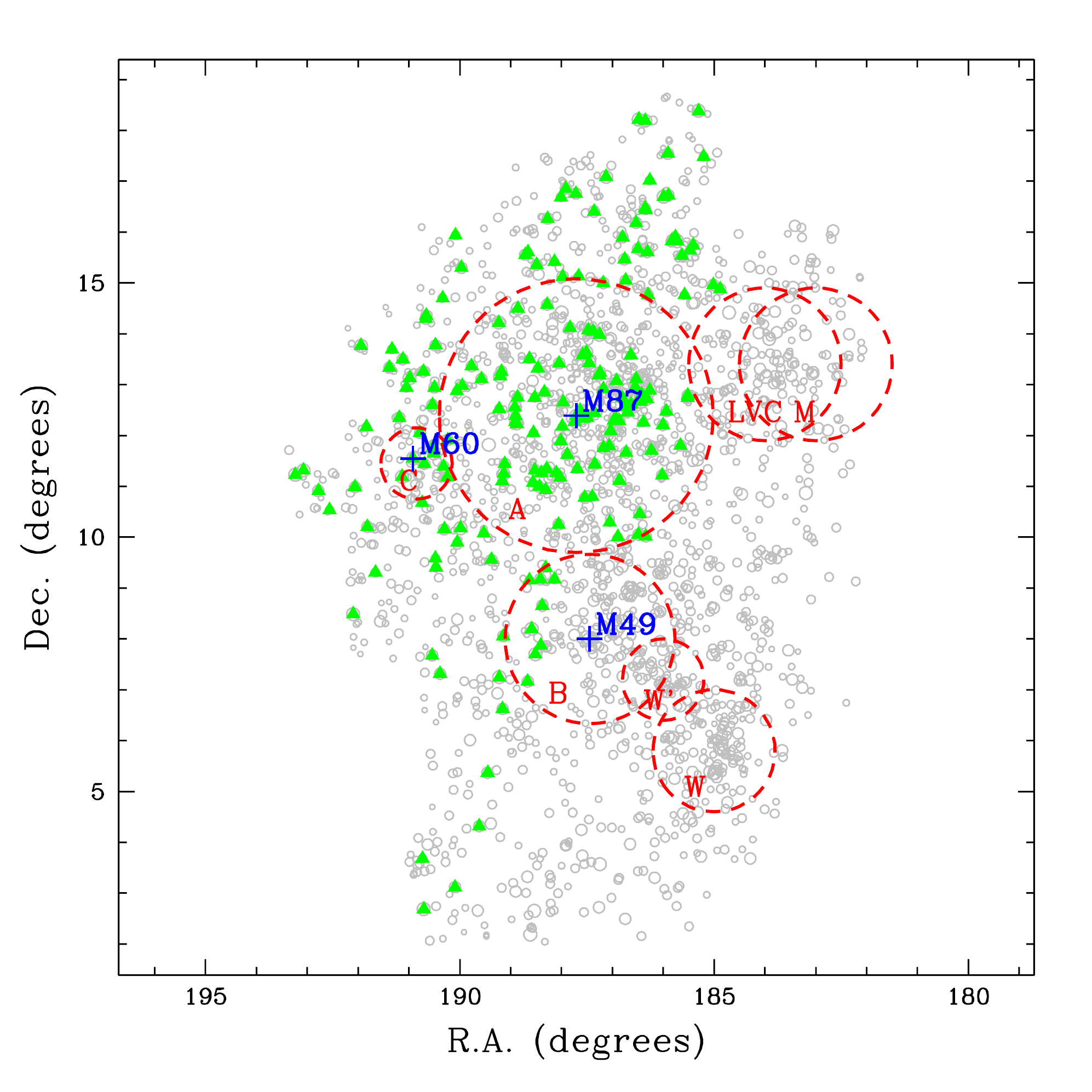}
    \caption{Overview of the Virgo Cluster, showing the main subcomponents. Open grey circles are galaxies from the VCC catalogue, with symbols size indicating $B_T$ magnitudes. 
Filled green triangles are objects in the SHIVir survey, which covers the eastern half of the cluster. Dashed red circles indicate the main substructures in the cluster, as defined in Boselli et al.~(2014).
Blue crosses indicate the positions of the central galaxies in components A, B, C.}
    \label{fig:6}
    \vspace{-0.48cm}
\end{figure}

The Virgo cluster is a relatively unrelaxed system, with clear spatial and kinematic substructure in the galaxy distribution \citep[e.g.][]{deVaucouleurs61,VCC,Gavazzi99, Boselli14}
that is also detected in X-ray emission \citep{Bohringer94} and the distribution of globular clusters \citep{Durrell14}. In particular, the cluster includes two main structural components, Virgo `A' centred on M87 and Virgo `B' centred on M49, several smaller subunits, including `Virgo C' centred on M60, a possible subunit centred on M86, the diffuse clouds W, W$^\prime$, and M, and the southern extension which leads away from the cluster to the southeast \citep{Boselli14}. Of these, the SHIVir survey covers components A and C, and the eastern edge of B, while avoiding most of the confusion from the background W, W$^\prime$, and M structures \citep{Ouellette17, Kourkchi17}.
Fig.~\ref{fig:6} illustrates the galaxy distribution from the VCC and SHIVir catalogues, as well as a few of the main subcomponents and their central galaxies. 

Given this structural complexity, we will consider two main indicators of a galaxy's environment. The first is the projected separation between the galaxy and M87, $R_{\rm p, M87}$, calculated in kpc or Mpc assuming both galaxies lie at the same line of sight distance, 16.5 Mpc \citep{Mei07,NGVS}. This measure has the advantage of simplicity, but ignores the substructure within the cluster. The second is $R_{\rm p, min}$, the projected separation from the centre of the nearest structural unit, scaled to the approximate size of that unit. Following \cite{Boselli14}, we take as units the components A, B and C, with centres at the positions of M87, M49 and M60 (i.e.~at coordinates (r.a., dec.) = (187.70$^\circ$, 12.39$^\circ$), (187.44$^\circ$, 8.00$^\circ$), (190.91$^\circ$, 11.55$^\circ$)) and sizes 2.692$^\circ$, 1.667$^\circ$, 0.7$^\circ$ (or $R_{\rm A},  R_{\rm B}, R_{\rm C}=$775, 480 and 202 kpc) respectively. 
The scaling with size reflects the fact that if components A, B and C are distinct, unmerged halos, with different masses and virial radii, 
we expect environmental trends to scale with virial radius, which we assume is proportional to the size defined in \cite{Boselli14}.  

\subsection{Density versus Environment: Basic Pattern} 
\label{subsec:4.2}

\begin{figure}
	\includegraphics[width=\columnwidth]{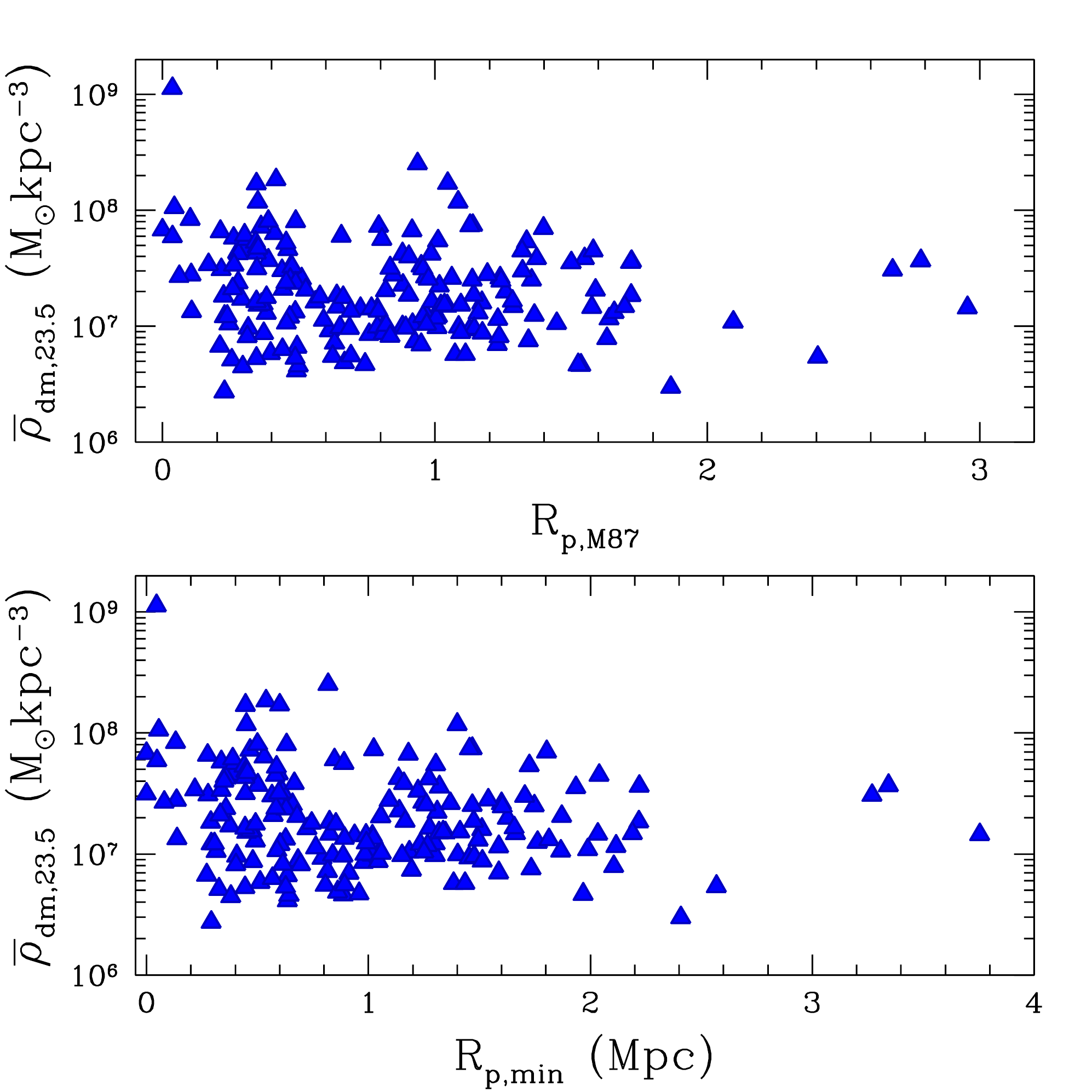}    
	\caption{Mean dark matter density within $r_{\rm o}$ versus { projected} position within the cluster, in terms of either $R_{p,M87}$ (top panel, in Mpc), or  $R_{\rm p, min}$ (bottom panel, in units of the size of the closest substructure, as defined in Boselli et al.~2014).}
    \label{fig:7}
\end{figure}

Fig.~\ref{fig:7} shows mean dark matter density within $r_{\rm o}$ versus position within the cluster, in terms of either $R_{p,M87}$ (top panel, in Mpc), or  $R_{\rm p, min}$ (bottom panel, in units of the size of the closest substructure, as defined in \cite{Boselli14}). In terms of  $R_{p,M87}$, we see a trend in density with projected separation, although it is obscured by a number of outliers. At separations of $<800$ kpc, approximately the size of region A, the trend is fairly strong; about half the galaxies in the range $R_{p,M87} = 0$--0.65 have densities $\bar{\rho}_{dm,23.5} > 3\times10^7$ M$_\odot$ kpc$^{-3}$, while only a few do in the range $R_{p,M87} = 0.65$--0.8. At larger separations, the median density is still fairly low, but there is more scatter. Some of this may reflect the complex structure of Virgo, however. In the lower panel, we plot density versus scaled distance to the nearest substructure,  $R_{\rm p, min}$. This shifts some of the denser galaxies to smaller separations, though the effect is not dramatic, possibly because the SHIVir survey only samples the structures A and C, and part of a third structure, B, so for most objects in the survey $R_{p,min}$ is proportional to $R_{p,M87}$. {The simulations show similar trends in density with environment, as discussed in Appendix \ref{appendix:simgradient}.}

\begin{figure*}[h]
	\includegraphics[width=1.99\columnwidth]{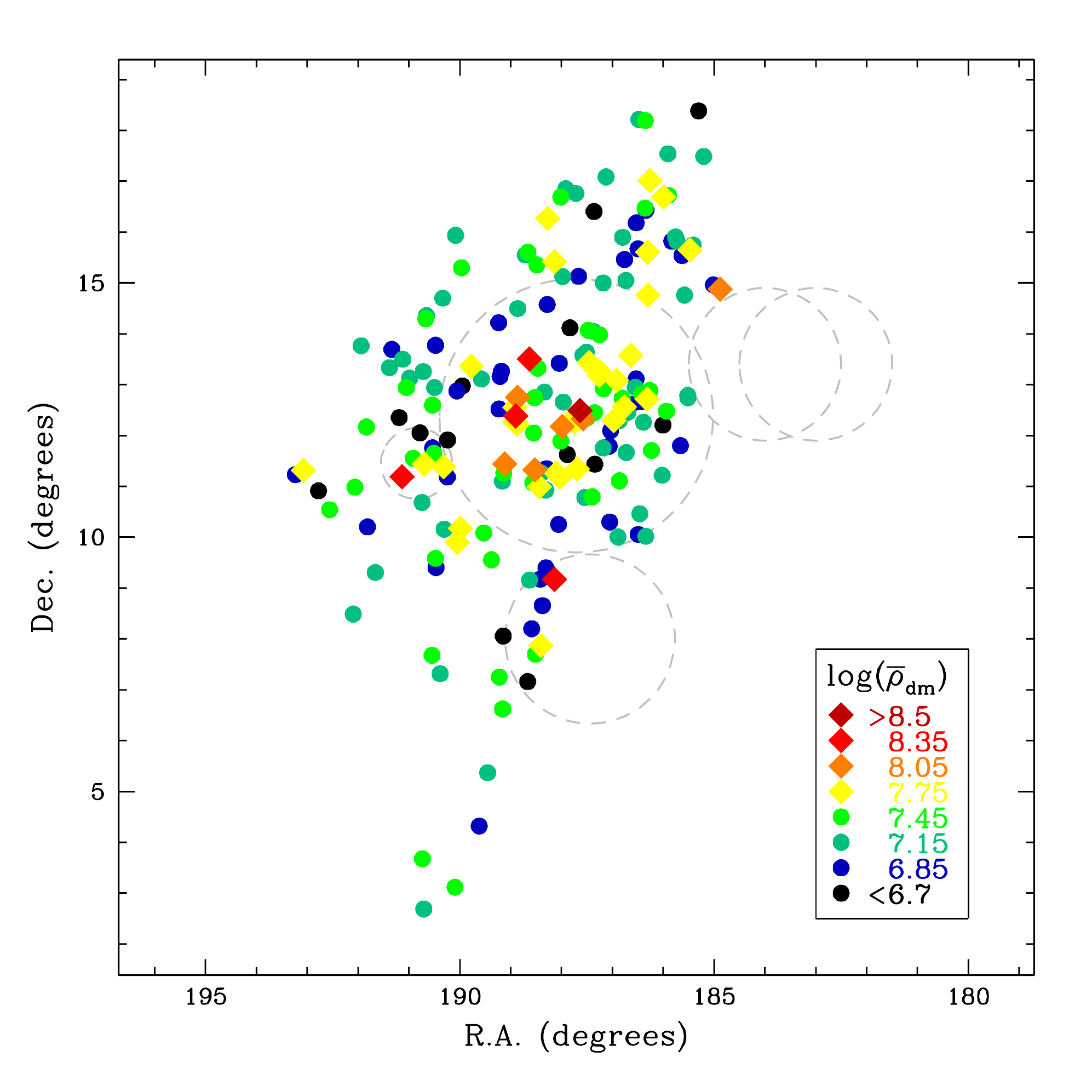}
    \caption{Density as a function of projected position for the SHIVir sample. { Symbol type and colour indicate the value of $\log(\bar{\rho}_{dm,23.5})$, with the binning indicated in units of $M_\odot\, {\rm kpc}^{-3}$.}
    Dashed circles indicate the positions and sizes of the substructures A, B, C, LVC and M discussed in section \ref{subsec:4.1}.}
    \label{fig:8}
\end{figure*}

We can get a better sense of environment by plotting dark matter density as a function of projected 2D position within the cluster. Fig.~\ref{fig:8} shows the distribution of the SHIVir sample, coloured by density. Symbol type indicates the value of $\log(\bar{\rho}_{dm,23.5})$, with black, blue, cyan and green circles indicating the ranges $<6.7$, 6.7--7.0, 7.0--7.3, and 7.3--7.6, 
while yellow, orange, red, and dark red diamonds indicate the ranges 7.6--7.9, 7.9--8.2, 8.2--8.5, and $>8.5$. Dashed circles indicate the positions and sizes of the substructures A, B, C, LVC and M discussed in section \ref{subsec:4.1}. In this view, the structure of the cluster emerges dramatically. Dense galaxies in and around region A show a clear spatial gradient ({ and the densest galaxy, VCC1297, lies very close to the centre of this region}). Outside this region, there are some dense galaxies, but some are clearly associated with component C, while others appear to lie with component B (although SHIVir coverage is very incomplete in this region). Despite the limited sample size, spatial incompleteness, and projection effects, there is a distinct impression that  the distribution of {$\bar{\rho}_{dm,23.5}$} is strongly correlated with local environment, with denser galaxies tracing out the cores of the oldest and densest subunits of the cluster. We will proceed to test this hypothesis statistically in the next section.
 
\subsection{Density versus Environment: Statistical Tests} 
\label{subsec:4.3}

There are several ways to test the significance of the pattern in Fig.~\ref{fig:8}. The simplest is to consider samples at large and small radii, and use a (two-sample) Kolmogorov-Smirnoff (KS) test to calculate the probability that either is drawn from the distribution given by the other. A free parameter in this test is the radius $R_{p,M87}$ at which to make the cut between the inner and outer 
samples. Given the pattern in Fig.~\ref{fig:8}, it seems this should be about half to two thirds the radius of component A. Comparing inner and out subsamples defined by various radial boundaries, we find  2.5-2.7\,$\sigma$ inconsistency between them ($p$-values of 0.007-0.012) over the range of cuts ($R_{p,M87}$ = 0.4--0.7 Mpc), that produce reasonable numbers of objects in both bins. The significance is slightly higher (SNR > 3) for cuts at very small radii ($R_{p,M87}$ < 0.2 Mpc), but we consider this result less robust given the small number of objects (< 10, or 5\%\ of the sample) in the inner bin. 

\begin{figure}
	\vspace{-2.5cm}
	\includegraphics[width=\columnwidth]{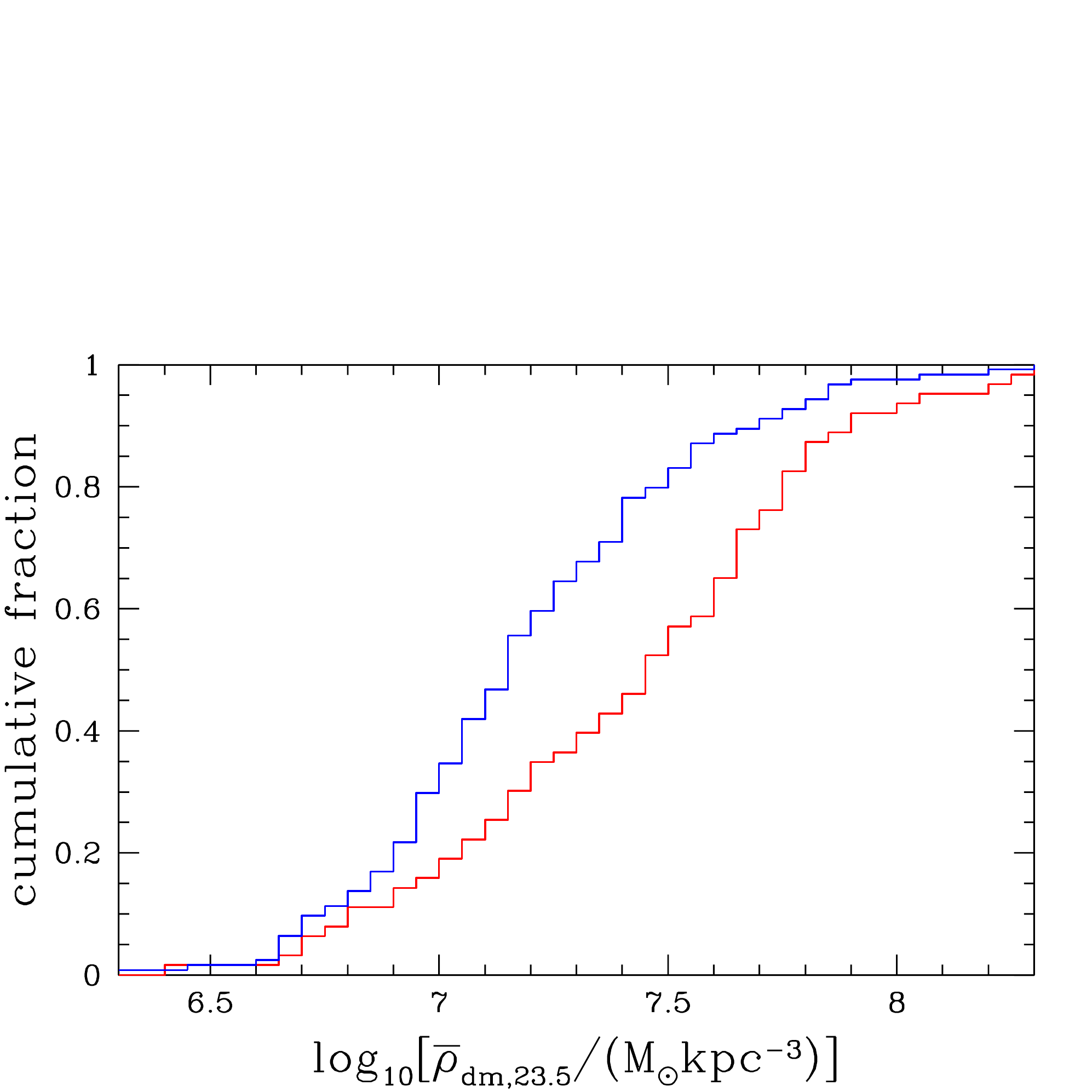}
    \caption{Cumulative distributions of dark matter density $\log[\bar{\rho}_{dm,23.5}/(M_\odot {\rm kpc}^{-3})]$, for samples with $R_{\rm p, min} \le 0.6$ and $R_{\rm p, min} > 0.6$.}
    \label{fig:9}
\end{figure}

{ By} selecting on $R_{p,M87}$ alone, we are ignoring the substructure around M60 and M49. If we select only the core region around M87, defined by $R_{p,M87} < R_{\rm A}$, we find that cuts around $R_{p,M87} \sim$ 0.5--0.6 Mpc show a discrepancy between the inner and outer samples at a significance of 2.8--3.1\,$\sigma$, while considering the whole cluster, but excluding regions close to M60 or M49 (objects with projected separations from these galaxies of less than 1.2 $R_{\rm C}$ or 1.2 $R_{\rm A}$ respectively), we find a significance of 2.6--3.2\,$\sigma$. We can include the clustering signal from galaxies in the secondary structural units by cutting on $R_{\rm p, min}$, the scaled distance to the nearest substructure. With this larger sample, for cuts  at $R_{\rm p, min}$ = 0.6--0.7, we now find a significance of 3.5--4\,$\sigma$. Figure \ref{fig:9} shows the normalized, cumulative distribution of  densities in inner and outer samples, cut at $R_{\rm p, min}$ = 0.6. The outer sample has a median density around the value $\log[\bar{\rho}_{dm,23.5}/(M_\odot {\rm kpc}^{-3})] = 7.15$, and few objects above 7.6. The inner sample has a 
median density around 7.45; it includes objects at low densities (some of which may be foreground/background systems with small projected separations), but a third of the objects lie at values of 7.6 or more. 

Given the trends in density with morphological type shown in Fig.~\ref{fig:5}, { the clustering signal could simply be} due to the morphology-density relation. We can test this possibility by splitting the sample by morphological type. { For subsamples that are large enough, we find that the clustering signal is still present}, albeit at slightly reduced significance. Thus, for instance cutting at $R_{\rm p, min}$ = 0.6, we find that early types (T < 3) show a signal at 2.8\,$\sigma$ significance, while late types (T = 3--13) show a signal at 2.3\,$\sigma$. For our other tests, the significance is generally in the range 2--3 sigma, and the cuts that produce the clearest separation between inner and outer samples are fairly consistent from one morphological type to the other. We conclude that the radial trends we see are robust to morphological selection, and are not simply a consequence of the morphology-density relation. (We have also considered the effect of stellar mass, and found that a clustering signal is present at both high and low stellar masses.)

The strength of the radial clustering signal is particularly striking, given possible contamination due to projection effects. We have experimented with cuts in distance, to select only those objects most likely to be close to the line-of-sight distance of the cluster. We find these are not very effective, removing objects from the inner and outer samples without increasing the significance of the discrepancy between them significantly. This may be because the typical distance errors are comparable to or larger than the size of the central region, and thus do not distinguish between projected and central objects very effectively. Redshift cuts are slightly more effective. In particular, restricting the sample to objects with $z = $0--0.008, we find a signal at a significance of 3.8-4.1\,$\sigma$ for  $R_{\rm p, min}$ = 0.6--0.7. Here too, though, the complex, broad velocity distribution within the cluster and its low redshift combine to make more accurate line-of-sight selection difficult. 

We have also tried several other two-sample KS tests, including tests on the distributions of $R_{\rm p, min}$ and $R_{p,M87}$, selected by $\bar{\rho}_{dm,23.5}$. We find a consistent pattern, albeit at comparable or slightly lower significance; dividing into high and low density samples above/below $\log[\bar{\rho}_{dm,23.5}/(M_\odot {\rm kpc}^{-3})] = 7.6$, for instance, we find a 3.5\,$\sigma$ discrepancy in the distribution of $R_{\rm p, min}$. Cutting the samples into high and low density at $\log[\bar{\rho}_{dm,23.5}/(M_\odot {\rm kpc}^{-3})] = 7.4$, 23\% of the low-density galaxies lie at $R_{\rm p, min} \le 0.6$, whereas 50\% of the high-density galaxies do. Finally, we have tried using the offset from the mean scaling relation  
$$\log[\bar{\rho}_{dm,23.5}/(M_\odot {\rm kpc}^{-3})] = 7.65 - 0.25\log(M_*/10^8 M_\odot)\,,$$ 
rather than the density $\bar{\rho}_{dm,23.5}$, in our previous tests. Here too, there is a clustering signal, but with a slightly lower (2.9\,$\sigma$) significance.

In summary, { the basic pattern seen in Fig.~\ref{fig:8}, that denser objects are preferentially located close to the centres of structures A, B and C, is found to hold at the 2--3\,$\sigma$ level for early or late morphological types considered separately. The significance reaches 3.5--4\,$\sigma$ when the entire sample is considered.} Although dark matter density does show a complex correlation with morphological type and stellar mass (cf.~Fig.~\ref{fig:5}), our tests with morphological or mass-selected subsamples indicate that { the trend in dark matter density with environment that we have identified is real and} spans a broad range of galaxy types and masses. 

\section{Discussion}
\label{sec:5}

\subsection{Comparison to Previous Work}
\label{subsec:5.1}

While the SHIVir survey provides contiguous coverage for a large fraction of the Virgo cluster, estimates of dark matter density are available for a number of other samples of
individual galaxies. 
\cite{Thomas09} derived mass models for 16 galaxies 
in the Coma cluster, while \cite{Alabi17} derived mass models for 32 galaxies in various environments. Enclosed dark matter masses and densities were calculated using mass modelling similar to the SHIVir analysis, but with apertures of $2 R_{\rm e}$ and $5 R_{\rm e}$ for the two studies, respectively. The former are generally smaller than our $R_{23.5}$, while the latter are smaller for $\sim 40$\%\ of the sample, so we correct all of these measurements to a common radius by dividing the ratio $g(x_{\rm ap})/g(x_{\rm o})$, where $g(x) = f(x)/x^{3}$ describes the scaling of the mean density with radius for an NFW profile. This correction requires a concentration parameter; we have calculated it assuming the mean concentration predicted by our analytic model for a galaxy of a given stellar mass at redshift $z=0$, and also using the concentrations determined for the individual systems in the case of the Alabi et al. sample. We find the differences between the resulting distributions are minor, so we will plot the former for simplicity.

\begin{figure}
	\vspace{-2cm}
	\includegraphics[width=\columnwidth]{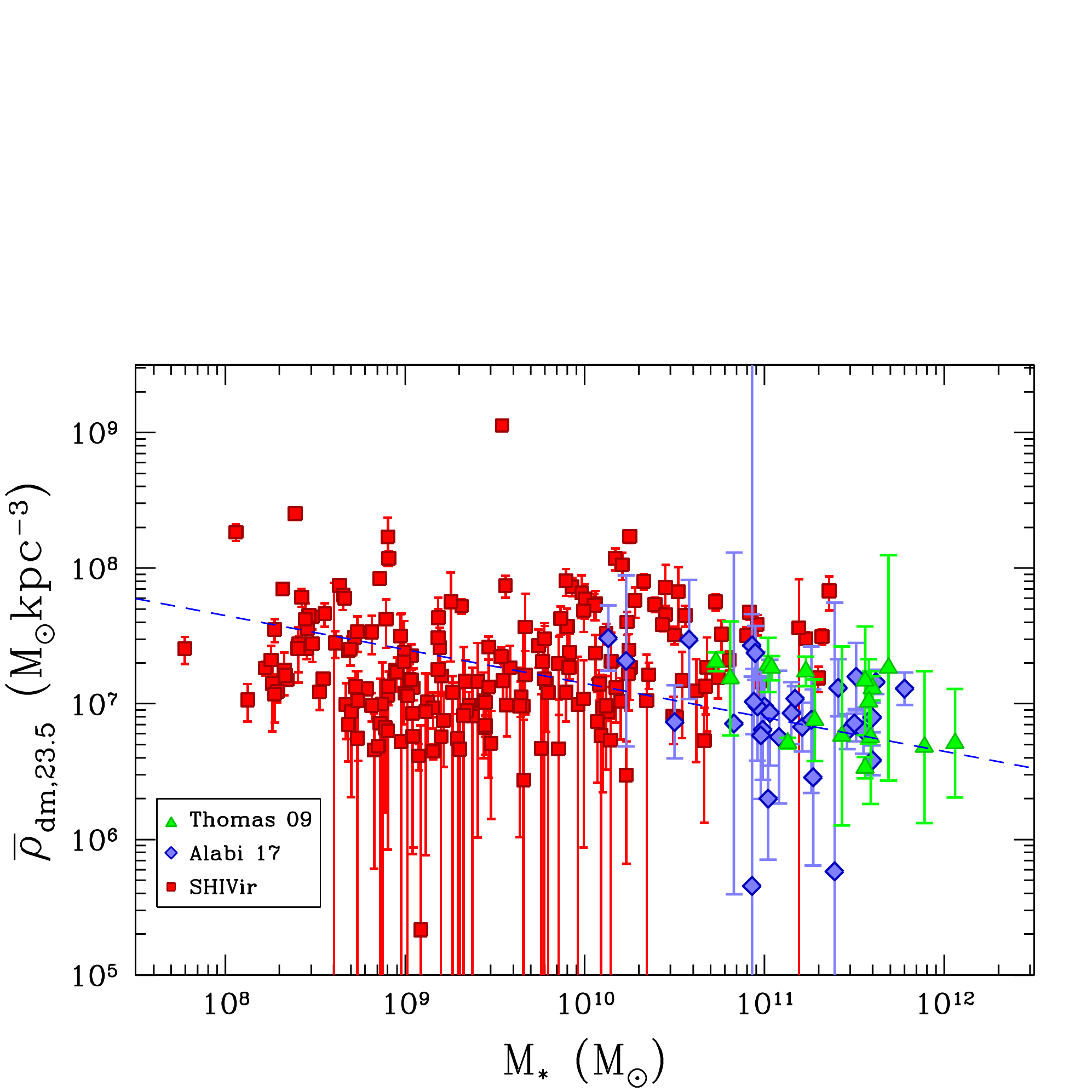}
    \caption{Mean dark matter density within $r_{\rm o}$ versus stellar mass, as in Fig.~\ref{fig:4}. Solid (red) squares are the SHIVir sample;  open (orange) squares are the sample of Alabi et al. (2017), from a range of environments, while open (green) triangles are the sample of Thomas et al. (2009) from the Coma cluster. The latter two data sets have both been corrected to a common radius $r_{\rm o}$, assuming dark matter follows an NFW profile with the concentration predicted by our analytic model at $z=0$. The dashed (blue) line shows the mean scaling relation $\bar{\rho}_{23.5} \propto M_*^{-0.25}$ derived for the SHIVir survey. }
    \label{fig:10}
\end{figure}

Fig.~\ref{fig:10} shows the samples from \cite{Thomas09} and \cite{Alabi17}, together with the SHIVir sample. We see that there is excellent agreement between the samples in the mean and scatter in dark matter density, and also in the slight decreasing trend with mass. The two other samples extend the SHIVir results to higher mass, but are generally consistent with the mean scaling relation $\bar{\rho}_{23.5} \propto M_*^{-0.25}$ derived for the SHIVir survey. The SHIVir data indicates that this fairly shallow relation extends over a surprising four orders of magnitude in mass. Considering the high mass end of the relation in more detail, we note the densities measured in the two other surveys lie at the lower end of the range seen in SHIVir. In deriving the correction to a common radius, we assumed $z=0$ concentrations; if we assumed higher redshift concentrations this would raise the points slightly. 
It may also be however, that the two other samples lie in local environments with lower densities. We will consider the properties of these other samples, notably the positions of the individual galaxies within groups or clusters, in more detail in future work.

\subsection{The Effect of Baryons}
\label{subsec:5.3}

So far in our interpretation of dark matter densities, we have ignored the complicating effects of baryons. Gas can cool and dissipate within dark matter halos, sinking to the centre and increasing the depth of the potential. If it occurs slowly, this process will drag the dark matter with it, increasing the dark matter density. The classic calculation of this effect (\citealt{Blumenthal86}; see also  \citealt{Eggen62, Barnes84, Ryden87}) assumes the contraction is slow enough that the adiabatic invariants of individual particle orbits are conserved. If these orbits are circular and shells of matter do not cross, the adiabatic invariant for a given shell of radius $r$ enclosing mass $M(r)$ is simply $rM(r)$, leading to the result
\begin{equation}
r_f/r_i = M_i/M_f\,.
\end{equation}
where subscripts $i$ and $f$ indicate the initial and final values, respectively.
\cite{Gnedin04} generalized this result to eccentric orbits, predicting a somewhat weaker contraction, as did \cite{Abadi10}. 
A convenient parameterization of the whole set of predictions, suggested by \cite{Dutton15}, is:
\begin{equation}
r_f/r_i = a + b(M_i/M_f + c)^d
\end{equation}
where $a$ = $c$ = 0, $b$ = $d$ = 1 for the classic prediction, and $a$ = $c$ = 0, $b = 1$, $d\sim0.8$ for the model of \cite{Gnedin04}.

{ Here}, we are typically measuring the mass distribution out at radii where dark matter dominates over baryons, so we expect even the classic estimate of adiabatic contraction to predict only modest changes in density. In Appendix \ref{appendix:contraction}, we derive the contraction factor for a region with a final stellar-to-dark-matter ratio $R_{bd}$: 
\begin{equation}
r_i/r_f = (R_{bd} + 1)(1-f_b)\, ,
\end{equation}
in the classic case, or
\begin{equation}
r_i/r_f = [(R_{bd} + 1)(1-f_b)]^{0.8}\, ,
\end{equation}
for the model of \cite{Gnedin04}, where $f_b$ is the baryon fraction of the initial halo before contraction.

The mean value of $R_{bd}$ at $r_{\rm o}$ for the SHIVir sample is 0.3, so assuming $f_b = \Omega_m/\Omega_{dm} = 0.186$, 
the contraction factor is $r_i/r_f = (R_{bd} + 1)(1-f_b) = 1.058$ for the classic calculation, or $r_i/r_f = [(R_{bd} + 1)(1-f_b)]^{0.8} =  1.046$ for the model of \cite{Gnedin04}.
At small radii, the enclosed dark matter mass goes roughly as $r^2$, and thus the dark matter mass contained within $r_{\rm o}$ will be $(r_i/r_f)^2 = 1.09$--1.12 times larger than expected from the initial halo profile, depending on the contraction model assumed. We conclude that while the dark matter densities measured in SHIVir may be slightly higher due to adiabatic contraction, the net effect is likely to be small for typical objects in our sample, even if we use the classical estimate. 

There are, however, a number of galaxies in the SHIVir sample with $R_{bd} > 1$, and even some that show little or no evidence for dark matter within $r_{\rm o}$. In these cases, baryons could have had a much larger relative effect, and the dark matter masses we measure could be much larger than the uncontracted values. Reducing the initial dark matter densities would move them further from the theoretical predictions, however, even for field objects forming at $z=0$, so it seems unlikely these objects have contracted significantly. One problem may be the large effect that systematics in the stellar masses have on the inferred dark matter density when $R_{bd}$ becomes small. There is also growing evidence, however, that baryonic effects may not cause contraction in all galaxies, but may have no effect, or even produce expansion in some cases.

There are several proposed mechanisms by which baryons may cause the potential to expand, including dynamical friction off infalling blobs of material, rapid, cyclic stellar feedback, supernova feedback, AGN feedback, or galactic bars \citep[see][for a recent review]{Dutton16}. Simulations suggest that the net effect on the potential may vary significantly from system to system \citep{Tissera10, Dutton16, Tollet16}, but with an overall dependence on the stellar-to-halo-mass ratio, and possibly on the size of the stellar system \citep{Dutton16}. Observationally, there are indications that the net results also depend on environment \citep[e.g.][]{Buote17}. We note, however, that almost all of these studies focus on effects at smaller radii; around $r_{\rm o}$, the net effect of baryonic processes, whether causing contraction or expansion, is small. 

{ In summary, current simulations and theoretical arguments do not uniquely identify how baryonic effects will influence the mass densities measured in our sample. It is clear, however, that these effects} will be small for most of the sample, which has $R_{bd} < 1$. For galaxies with low dark matter fractions, the low densities inferred from mass modelling suggest that some net expansion may have taken place. These are mainly dwarf elliptical galaxies, with stellar masses in the range $M_* = 10^{8.5}$--$10^{9.5} M_\odot$. In that sense, SHIVir measurements are consistent with theoretical models suggesting that low-mass galaxies may have expanded due to violent feedback processes, and may now have cored density profiles \citep{DiCintio14}. 

{ Finally, we address any possible effect on the clustering signal.} We have rerun our statistical tests, excluding objects with very small dark matter fractions, and find that the clustering pattern remains robust, albeit at slightly reduced significance if we eliminate too many systems. Systems with high dark matter fractions within $r_{\rm o}$ are unlikely to be affected by adiabatic contraction, and thus their high dark matter densities are likely to be primordial. The status of the objects with very low dark matter densities is less clear; {  they may have experienced some expansion due to baryonic effects, but current mass modelling uncertainties do not allow firm conclusions on an object-by-object basis \citep{Courteau14}.}

\subsection{Connection to Fundamental Plane Offsets}
\label{subsec:5.4}

It is interesting to explore the connection between dark matter density and other structural properties of galaxies, such as the Fundamental Plane (FP), the correlation between size, surface brightness and velocity dispersion. \cite{Ouellette17} constructed the FP for the 88 early-type (E or dE to Sa) galaxies in Virgo, and examined residuals with respect to the plane, as a function of various galaxy properties. They found correlations with two related properties, the total-mass-to-light ratio and the stellar-to-total mass ratio; they considered the latter more significant and more fundamental.

\begin{figure}
	\includegraphics[width=\columnwidth]{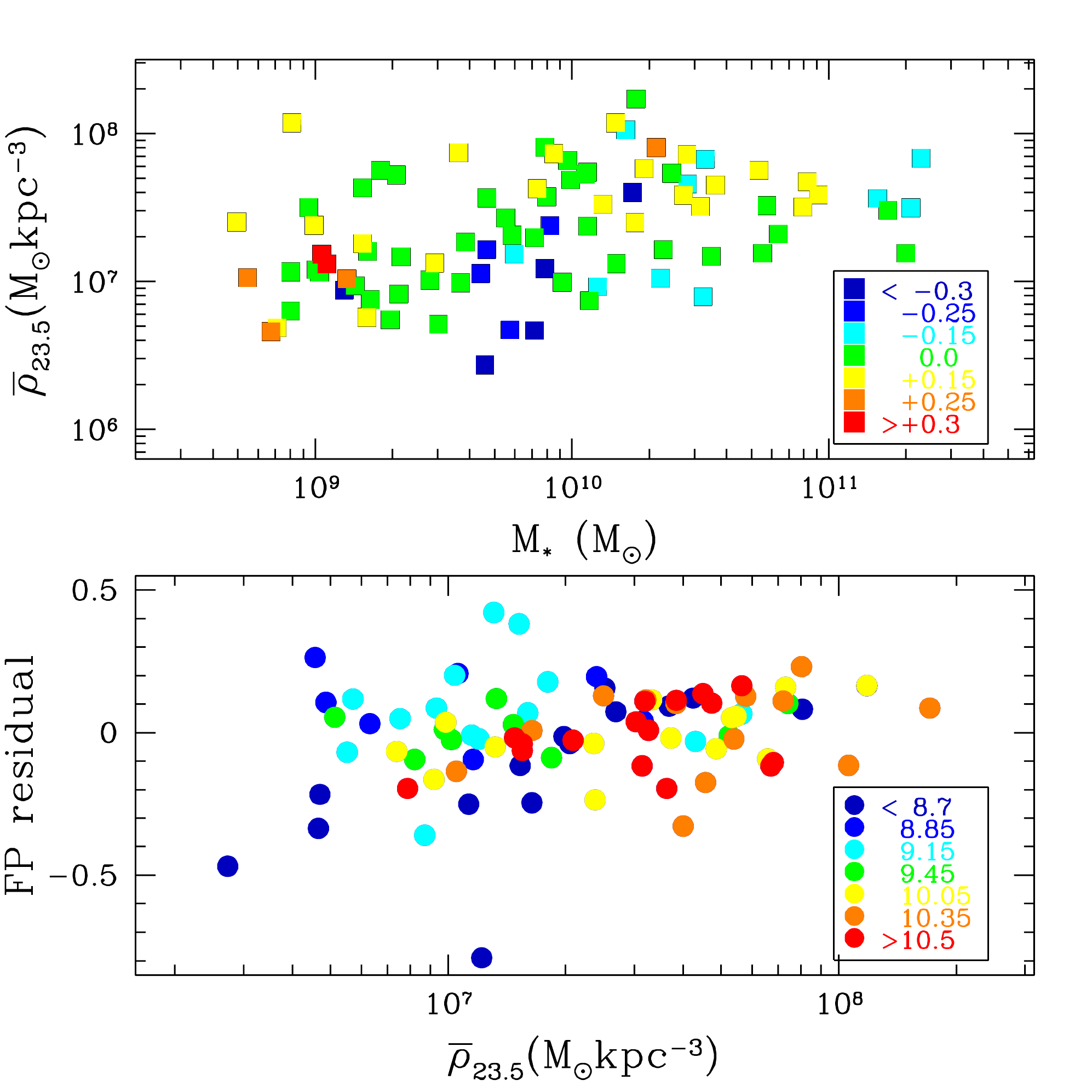}
    \caption{{\it Top panel:} Dark matter density versus stellar mass, with points coloured by FP residual, { with the binning indicated} {\it Bottom panel:} FP residuals versus dark matter density, with points coloured by stellar mass, { with the binning in $M_*/M_\odot$ indicated.}}
    \label{fig:11}
\end{figure}

In Fig.~\ref{fig:11}, we show dark matter density versus stellar mass, with points coloured by FP residual (top panel -- black/blue negative residuals, orange/red positive residuals), and FP residuals versus dark matter density, with points coloured by stellar mass (bottom panel -- black/blue low stellar masses, orange/red high stellar masses). Examining these two plots, we find several trends. First, FP residuals do not correlate monotonically with stellar mass nor with dark matter density. On the other hand, in this plane, objects with high FP residuals do seem to fall into two categories, either objects with large stellar masses and high dark matter densities, that we infer to be the oldest massive galaxies in the central region of the cluster, or the opposite extreme, low-mass galaxies with low densities, which could be dEs or dS0s affected by feedback or harassment. Between these two regimes, objects have small and/or negative FP residuals. 

In the bottom panel, we see a different projection of the same trend. In general, FP residuals increase with dark matter density, and the densest objects are high-stellar mass galaxies with positive FP residuals. On the other hand,  some low stellar mass systems have low dark matter densities and a large range of (mainly positive)  FP residuals. It seems likely that { multiple processes are contributing to the scatter in the FP, and more detailed work will be} required to establish the exact mechanisms involved. We have also examined FP residuals as a function of spatial location, but do not find any single clear pattern, unlike for dark matter density. 

\subsection{Outliers from the Mean Radial Trend and Kinematic Groups}
\label{subsec:5.5}

While the presence of low-density objects at small separations from the centres of structures A, B and C is easily understood as a consequence of projection effects, or recent infall from the field on very radial orbits, the presence of dense objects outside the cluster core is { less intuitive}. We have examined the dozen galaxies in Fig.~\ref{fig:8} with densities  $\log[\bar{\rho}_{dm,23.5}/(M_\odot {\rm kpc}^{-3})] > 7.6$ {that are located outside dense regions}, and find that they do not consists of a single stellar mass range or morphological type, nor do their images show obvious structural peculiarities (although they do generally have positive FP residuals). 
Individually, it is possible these objects formed at high redshift and have not grown much since, giving them high central dark matter densities. We note, however, that they appear to be slightly clustered on the sky, with 8 of them lying to the north-west of structure A. Given the presence of other substructure in Virgo in this region, it may be that these galaxies represent the core of a disrupted group falling into the main cluster. Unfortunately, the SHIV sample is too sparse, and the distance estimates to these galaxies are too imprecise, to reach a firm conclusion.  

Selecting objects by density and considering their projected positions and kinematics {reveals other evidence for dynamical} substructure in the cluster. In particular, many of the dense galaxies close to M87 appear to { form} a ring-like structure, and have smoothly varying velocities as a function of position. { This may well represent a disrupting group, and/or have some connection to the concentration of galaxies around M86 (west of M87), but here too the sampling is too coarse to reach firm conclusions}. 

\section{Conclusions}
\label{sec:6}

In this paper, we have considered the mean density of dark matter interior to the visible edge of a galaxy, $\bar{\rho}_{dm,23.5}$ (defined in section \ref{subsec:2.1}). Assuming the standard empirical relations between stellar mass and halo mass, halo mass and concentration, and stellar mass and galaxy size, we have shown that the mean dark matter density is expected to be approximately independent of halo mass, but to vary with the redshift $z_{\rm f}$ at which the halo `formed' or established its central core density. Using cosmological, dark-matter only simulations of cluster formation, we have shown that this central density is generally conserved when field halos merge into a larger group or cluster environment, except perhaps for the most massive objects. We conclude that dark matter density may provide a direct measure of formation epoch, and thus place limits on the `infall' redshift at which a galaxy first merged into a group or cluster. 

{ We have tested this hypothesis in the Virgo cluster, using data from the SHIVir survey. We find} that there is indeed a clear pattern in the projected spatial distribution of galaxies as a function of their dark matter density. The densest objects ($\log[\bar{\rho}_{dm,23.5}/(M_\odot {\rm kpc}^{-3})] > 7.4$) appear to trace out the core of the main structural component of the cluster, centred on M87, as well as the subcomponents centred on M60 and M49. Statistical (KS) tests indicate the distributions of density in inner and outer samples are discrepant at surprisingly high (3.5--4\,$\sigma$) significance, given projection effects, the complex structure of the cluster, and the limited sample size. These results { confirm previously found trends in density with environment \citep{Thomas09,Alabi17},
but provide much more complete coverage in stellar mass and area on the sky within a single cluster, demonstrating for the first time the full potential of this test.}

Clearly, many galaxy properties vary systematically between the core and outskirts of a cluster, including mean stellar mass, stellar age, star formation history, colour and morphology. 
We emphasize that dark matter density is different, since it is thought to be primarily determined by the assembly history of the halo within which the galaxy first formed. In particular, the formation time of the halo is distinct from the mean stellar age, as emphasized by \cite{Hill17}; 
 stars may form in a set of smaller galaxies that assemble in a dry merger,  producing a stellar age older than the halo formation time, or star formation may be triggered after a halo assembles, producing a stellar age younger than the halo age. Knowing the formation epoch of a halo and its current position gives a much better constraint on its original (field) mass before it fell into the cluster, and thus allows for a much more precise version of (substructure) abundance matching \cite[e.g.][]{Grossauer15}. 

The interpretation of $\bar{\rho}_{dm,23.5}$ as an indicator of halo formation time { carries a few caveats}. The first and most important is the influence of baryons on the present-day dark matter density. The classic estimate of adiabatic contraction suggests that it is unimportant for most { galaxies in our sample; given they are generally dark-matter dominated within $R_{23.5}$, the predicted increase in density is 30\% or less.} Some of the objects with the lowest dark matter densities { may} have experienced expansion due to violent feedback, but our tests indicate that this does not { affect the overall correlation of $\bar{\rho}_{dm,23.5}$ with environment significantly}. A second, more minor concern are the systematics involved in mass modelling, including the IMF, star formation histories and distances assumed. These may introduce stellar mass errors significantly larger than the ones we have assumed \citep{Courteau14}. On the other hand, we would expect larger errors to wash out { our} inferred clustering signal, so we conclude that either these errors are not { significantly} larger than assumed, or the underlying clustering signal is extremely strong. These additional systematics are most concerning when comparing results from different surveys and in different environments, and suggest the need for large, homogeneous studies using a single mass-modelling method. 

Overall, measurements of dark matter density appear to provide a sensitive observational test of halo formation times and infall redshifts. { We have demonstrated the potential of this method with a large, homogeneous sample of mass estimates from the SHIVir survey of Virgo cluster galaxies. The Virgo cluster is a complex environment, however, with multiple structural components.} Similar studies in other environments such as the Coma or Fornax clusters could easily confirm the patterns seen in this work (and also suggested previously by studies with smaller samples -- \citealt{Thomas09,Alabi17}). With the advent of  new spectroscopic facilities, notably integral field units \citep[e.g.][]{CALIFA,SAMI,MANGA},
there are excellent prospects for mapping out the density of nearby galaxies across a wide range of environments. 

\section*{Acknowledgements}

JET { and SC acknowledge support from the Natural Sciences and Engineering Research Council (NSERC) of Canada through respective Discovery Grants. NNQO acknowledges support from NSERC through a PGS D scholarship. We thank the anonymous referee for several comments and questions that improved the paper significantly. We also thank B. Tully, M. Hudson, N. Dalal, and the participants of the conference ``The Physics of Galaxy Scaling Relations and the Nature of Dark Matter'' held at Queen's University in July 2018} for many useful comments on a preliminary version of this work. 
This work made use of the VizieR database (https://vizier.u-strasbg.fr/vizier).


\bibliographystyle{mnras}
\bibliography{assembly_virgo} 

\begin{thebibliography}{}
\makeatletter
\relax
\def\mn@urlcharsother{\let\do\@makeother \do\$\do\&\do\#\do\^\do\_\do\%\do\~}
\def\mn@doi{\begingroup\mn@urlcharsother \@ifnextchar [ {\mn@doi@}
  {\mn@doi@[]}}
\def\mn@doi@[#1]#2{\def\@tempa{#1}\ifx\@tempa\@empty \href
  {http://dx.doi.org/#2} {doi:#2}\else \href {http://dx.doi.org/#2} {#1}\fi
  \endgroup}
\def\mn@eprint#1#2{\mn@eprint@#1:#2::\@nil}
\def\mn@eprint@arXiv#1{\href {http://arxiv.org/abs/#1} {{\tt arXiv:#1}}}
\def\mn@eprint@dblp#1{\href {http://dblp.uni-trier.de/rec/bibtex/#1.xml}
  {dblp:#1}}
\def\mn@eprint@#1:#2:#3:#4\@nil{\def\@tempa {#1}\def\@tempb {#2}\def\@tempc
  {#3}\ifx \@tempc \@empty \let \@tempc \@tempb \let \@tempb \@tempa \fi \ifx
  \@tempb \@empty \def\@tempb {arXiv}\fi \@ifundefined
  {mn@eprint@\@tempb}{\@tempb:\@tempc}{\expandafter \expandafter \csname
  mn@eprint@\@tempb\endcsname \expandafter{\@tempc}}}

\bibitem[\protect\citeauthoryear{{Abadi}, {Navarro}, {Fardal}, {Babul}  \&
  {Steinmetz}}{{Abadi} et~al.}{2010}]{Abadi10}
{Abadi} M.~G.,  {Navarro} J.~F.,  {Fardal} M.,  {Babul} A.,   {Steinmetz} M.,
  2010, \mn@doi [\mnras] {10.1111/j.1365-2966.2010.16912.x}, \href
  {http://adsabs.harvard.edu/abs/2010MNRAS.407..435A} {407, 435}

\bibitem[\protect\citeauthoryear{{Adelman-McCarthy} et~al.,}{{Adelman-McCarthy}
  et~al.}{2008}]{SDSS6}
{Adelman-McCarthy} J.~K.,  et~al., 2008, \mn@doi [\apjs] {10.1086/524984},
  \href {http://adsabs.harvard.edu/abs/2008ApJS..175..297A} {175, 297}

\bibitem[\protect\citeauthoryear{{Alabi} et~al.,}{{Alabi}
  et~al.}{2017}]{Alabi17}
{Alabi} A.~B.,  et~al., 2017, \mn@doi [\mnras] {10.1093/mnras/stx678}, \href
  {http://adsabs.harvard.edu/abs/2017MNRAS.468.3949A} {468, 3949}

\bibitem[\protect\citeauthoryear{{Balogh} et~al.,}{{Balogh}
  et~al.}{2004a}]{Balogh04b}
{Balogh} M.,  et~al., 2004a, \mn@doi [\mnras]
  {10.1111/j.1365-2966.2004.07453.x}, \href
  {http://adsabs.harvard.edu/abs/2004MNRAS.348.1355B} {348, 1355}

\bibitem[\protect\citeauthoryear{{Balogh}, {Baldry}, {Nichol}, {Miller},
  {Bower}  \& {Glazebrook}}{{Balogh} et~al.}{2004b}]{Balogh04a}
{Balogh} M.~L.,  {Baldry} I.~K.,  {Nichol} R.,  {Miller} C.,  {Bower} R.,
  {Glazebrook} K.,  2004b, \mn@doi [\apjl] {10.1086/426079}, \href
  {http://adsabs.harvard.edu/abs/2004ApJ...615L.101B} {615, L101}

\bibitem[\protect\citeauthoryear{{Barnes} \& {White}}{{Barnes} \&
  {White}}{1984}]{Barnes84}
{Barnes} J.,  {White} S.~D.~M.,  1984, \mn@doi [\mnras]
  {10.1093/mnras/211.4.753}, \href
  {http://adsabs.harvard.edu/abs/1984MNRAS.211..753B} {211, 753}

\bibitem[\protect\citeauthoryear{{Behroozi}, {Conroy}  \&
  {Wechsler}}{{Behroozi} et~al.}{2010}]{Behroozi10}
{Behroozi} P.~S.,  {Conroy} C.,   {Wechsler} R.~H.,  2010, \mn@doi [\apj]
  {10.1088/0004-637X/717/1/379}, \href
  {http://adsabs.harvard.edu/abs/2010ApJ...717..379B} {717, 379}

\bibitem[\protect\citeauthoryear{{Behroozi}, {Wechsler}  \&
  {Conroy}}{{Behroozi} et~al.}{2013}]{Behroozi13}
{Behroozi} P.~S.,  {Wechsler} R.~H.,   {Conroy} C.,  2013, \mn@doi [\apj]
  {10.1088/0004-637X/770/1/57}, \href
  {http://adsabs.harvard.edu/abs/2013ApJ...770...57B} {770, 57}

\bibitem[\protect\citeauthoryear{{Binggeli}, {Sandage}  \&
  {Tammann}}{{Binggeli} et~al.}{1985a}]{Binggeli85}
{Binggeli} B.,  {Sandage} A.,   {Tammann} G.~A.,  1985a, \mn@doi [\aj]
  {10.1086/113874}, \href {http://adsabs.harvard.edu/abs/1985AJ.....90.1681B}
  {90, 1681}

\bibitem[\protect\citeauthoryear{{Binggeli}, {Sandage}  \&
  {Tammann}}{{Binggeli} et~al.}{1985b}]{VCC}
{Binggeli} B.,  {Sandage} A.,   {Tammann} G.~A.,  1985b, \mn@doi [\aj]
  {10.1086/113874}, \href {http://adsabs.harvard.edu/abs/1985AJ.....90.1681B}
  {90, 1681}

\bibitem[\protect\citeauthoryear{{Biviano}, {Katgert}, {Mazure}, {Moles}, {den
  Hartog}, {Perea}  \& {Focardi}}{{Biviano} et~al.}{1997}]{Biviano97}
{Biviano} A.,  {Katgert} P.,  {Mazure} A.,  {Moles} M.,  {den Hartog} R.,
  {Perea} J.,   {Focardi} P.,  1997, \aap, \href
  {http://adsabs.harvard.edu/abs/1997A%26A...321...84B} {321, 84}

\bibitem[\protect\citeauthoryear{{Blumenthal}, {Faber}, {Flores}  \&
  {Primack}}{{Blumenthal} et~al.}{1986}]{Blumenthal86}
{Blumenthal} G.~R.,  {Faber} S.~M.,  {Flores} R.,   {Primack} J.~R.,  1986,
  \mn@doi [\apj] {10.1086/163867}, \href
  {http://adsabs.harvard.edu/abs/1986ApJ...301...27B} {301, 27}

\bibitem[\protect\citeauthoryear{{B{\"o}hringer}, {Briel}, {Schwarz}, {Voges},
  {Hartner}  \& {Tr{\"u}mper}}{{B{\"o}hringer} et~al.}{1994}]{Bohringer94}
{B{\"o}hringer} H.,  {Briel} U.~G.,  {Schwarz} R.~A.,  {Voges} W.,  {Hartner}
  G.,   {Tr{\"u}mper} J.,  1994, \mn@doi [\nat] {10.1038/368828a0}, \href
  {http://adsabs.harvard.edu/abs/1994Natur.368..828B} {368, 828}

\bibitem[\protect\citeauthoryear{{Boselli} et~al.,}{{Boselli}
  et~al.}{2014}]{Boselli14}
{Boselli} A.,  et~al., 2014, \mn@doi [\aap] {10.1051/0004-6361/201424419},
  \href {http://adsabs.harvard.edu/abs/2014A\%26A...570A..69B} {570, A69}

\bibitem[\protect\citeauthoryear{{Bryant et al.}}{{Bryant et al.}}{2015}]{SAMI}
{Bryant et al.} J.~J.,  2015, \mn@doi [\mnras] {10.1093/mnras/stu2635}, \href
  {http://adsabs.harvard.edu/abs/2015MNRAS.447.2857B} {447, 2857}

\bibitem[\protect\citeauthoryear{{Bundy et al.}}{{Bundy et al.}}{2015}]{MANGA}
{Bundy et al.} K.,  2015, \mn@doi [\apj] {10.1088/0004-637X/798/1/7}, \href
  {http://adsabs.harvard.edu/abs/2015ApJ...798....7B} {798, 7}

\bibitem[\protect\citeauthoryear{{Buote}}{{Buote}}{2017}]{Buote17}
{Buote} D.~A.,  2017, \mn@doi [\apj] {10.3847/1538-4357/834/2/164}, \href
  {http://adsabs.harvard.edu/abs/2017ApJ...834..164B} {834, 164}

\bibitem[\protect\citeauthoryear{{C{\^o}t{\'e}} et~al.,}{{C{\^o}t{\'e}}
  et~al.}{2004}]{Cote04}
{C{\^o}t{\'e}} P.,  et~al., 2004, \mn@doi [\apjs] {10.1086/421490}, \href
  {http://adsabs.harvard.edu/abs/2004ApJS..153..223C} {153, 223}

\bibitem[\protect\citeauthoryear{{Courteau} et~al.,}{{Courteau}
  et~al.}{2014}]{Courteau14}
{Courteau} S.,  et~al., 2014, \mn@doi [Reviews of Modern Physics]
  {10.1103/RevModPhys.86.47}, \href
  {https://ui.adsabs.harvard.edu/abs/2014RvMP...86...47C} {86, 47}

\bibitem[\protect\citeauthoryear{{Di Cintio}, {Brook}, {Macci{\`o}}, {Stinson},
  {Knebe}, {Dutton}  \& {Wadsley}}{{Di Cintio} et~al.}{2014}]{DiCintio14}
{Di Cintio} A.,  {Brook} C.~B.,  {Macci{\`o}} A.~V.,  {Stinson} G.~S.,  {Knebe}
  A.,  {Dutton} A.~A.,   {Wadsley} J.,  2014, \mn@doi [\mnras]
  {10.1093/mnras/stt1891}, \href
  {http://adsabs.harvard.edu/abs/2014MNRAS.437..415D} {437, 415}

\bibitem[\protect\citeauthoryear{{Diemand}, {Moore}, {Stadel}  \&
  {Kazantzidis}}{{Diemand} et~al.}{2004}]{Diemand04}
{Diemand} J.,  {Moore} B.,  {Stadel} J.,   {Kazantzidis} S.,  2004, \mn@doi
  [\mnras] {10.1111/j.1365-2966.2004.07424.x}, \href
  {http://adsabs.harvard.edu/abs/2004MNRAS.348..977D} {348, 977}

\bibitem[\protect\citeauthoryear{{Diemer} \& {Kravtsov}}{{Diemer} \&
  {Kravtsov}}{2015}]{Diemer15}
{Diemer} B.,  {Kravtsov} A.~V.,  2015, \mn@doi [apj]
  {10.1088/0004-637X/799/1/108}, \href
  {http://adsabs.harvard.edu/abs/2015ApJ...799..108D} {799, 108}

\bibitem[\protect\citeauthoryear{{Drakos}, {Taylor}  \& {Benson}}{{Drakos}
  et~al.}{2017}]{Drakos17}
{Drakos} N.~E.,  {Taylor} J.~E.,   {Benson} A.~J.,  2017, \mn@doi [\mnras]
  {10.1093/mnras/stx652}, \href
  {http://adsabs.harvard.edu/abs/2017MNRAS.468.2345D} {468, 2345}

\bibitem[\protect\citeauthoryear{{Duffy}, {Schaye}, {Kay}  \& {Dalla
  Vecchia}}{{Duffy} et~al.}{2008}]{Duffy08}
{Duffy} A.~R.,  {Schaye} J.,  {Kay} S.~T.,   {Dalla Vecchia} C.,  2008, \mn@doi
  [mnras] {10.1111/j.1745-3933.2008.00537.x}, \href
  {http://adsabs.harvard.edu/abs/2008MNRAS.390L..64D} {390, L64}

\bibitem[\protect\citeauthoryear{{Durrell} et~al.,}{{Durrell}
  et~al.}{2014}]{Durrell14}
{Durrell} P.~R.,  et~al., 2014, \mn@doi [\apj] {10.1088/0004-637X/794/2/103},
  \href {http://adsabs.harvard.edu/abs/2014ApJ...794..103D} {794, 103}

\bibitem[\protect\citeauthoryear{{Dutton}, {Macci{\`o}}, {Stinson}, {Gutcke},
  {Penzo}  \& {Buck}}{{Dutton} et~al.}{2015}]{Dutton15}
{Dutton} A.~A.,  {Macci{\`o}} A.~V.,  {Stinson} G.~S.,  {Gutcke} T.~A.,
  {Penzo} C.,   {Buck} T.,  2015, \mn@doi [\mnras] {10.1093/mnras/stv1755},
  \href {http://adsabs.harvard.edu/abs/2015MNRAS.453.2447D} {453, 2447}

\bibitem[\protect\citeauthoryear{{Dutton} et~al.,}{{Dutton}
  et~al.}{2016}]{Dutton16}
{Dutton} A.~A.,  et~al., 2016, \mn@doi [\mnras] {10.1093/mnras/stw1537}, \href
  {http://adsabs.harvard.edu/abs/2016MNRAS.461.2658D} {461, 2658}

\bibitem[\protect\citeauthoryear{{Eggen}, {Lynden-Bell}  \& {Sandage}}{{Eggen}
  et~al.}{1962}]{Eggen62}
{Eggen} O.~J.,  {Lynden-Bell} D.,   {Sandage} A.~R.,  1962, \mn@doi [\apj]
  {10.1086/147433}, \href {http://adsabs.harvard.edu/abs/1962ApJ...136..748E}
  {136, 748}

\bibitem[\protect\citeauthoryear{{Einasto}}{{Einasto}}{1965}]{Einasto65}
{Einasto} J.,  1965, Trudy Astrofizicheskogo Instituta Alma-Ata, \href
  {http://adsabs.harvard.edu/abs/1965TrAlm...5...87E} {5, 87}

\bibitem[\protect\citeauthoryear{{Ellingson}, {Lin}, {Yee}  \&
  {Carlberg}}{{Ellingson} et~al.}{2001}]{Ellingson01}
{Ellingson} E.,  {Lin} H.,  {Yee} H.~K.~C.,   {Carlberg} R.~G.,  2001, \mn@doi
  [\apj] {10.1086/318423}, \href
  {http://adsabs.harvard.edu/abs/2001ApJ...547..609E} {547, 609}

\bibitem[\protect\citeauthoryear{{Faber} et~al.,}{{Faber}
  et~al.}{2007}]{Faber07}
{Faber} S.~M.,  et~al., 2007, \mn@doi [\apj] {10.1086/519294}, \href
  {http://adsabs.harvard.edu/abs/2007ApJ...665..265F} {665, 265}

\bibitem[\protect\citeauthoryear{{Ferrarese} et~al.,}{{Ferrarese}
  et~al.}{2012}]{NGVS}
{Ferrarese} L.,  et~al., 2012, \mn@doi [\apjs] {10.1088/0067-0049/200/1/4},
  \href {http://adsabs.harvard.edu/abs/2012ApJS..200....4F} {200, 4}

\bibitem[\protect\citeauthoryear{{Gao}, {Navarro}, {Cole}, {Frenk}, {White},
  {Springel}, {Jenkins}  \& {Neto}}{{Gao} et~al.}{2008}]{Gao08}
{Gao} L.,  {Navarro} J.~F.,  {Cole} S.,  {Frenk} C.~S.,  {White} S.~D.~M.,
  {Springel} V.,  {Jenkins} A.,   {Neto} A.~F.,  2008, \mn@doi [mnras]
  {10.1111/j.1365-2966.2008.13277.x}, \href
  {http://adsabs.harvard.edu/abs/2008MNRAS.387..536G} {387, 536}

\bibitem[\protect\citeauthoryear{{Gavazzi}, {Boselli}, {Scodeggio}, {Pierini}
  \& {Belsole}}{{Gavazzi} et~al.}{1999}]{Gavazzi99}
{Gavazzi} G.,  {Boselli} A.,  {Scodeggio} M.,  {Pierini} D.,   {Belsole} E.,
  1999, \mn@doi [\mnras] {10.1046/j.1365-8711.1999.02350.x}, \href
  {http://adsabs.harvard.edu/abs/1999MNRAS.304..595G} {304, 595}

\bibitem[\protect\citeauthoryear{{Gavazzi}, {Boselli}, {Donati}, {Franzetti}
  \& {Scodeggio}}{{Gavazzi} et~al.}{2003}]{Gavazzi03}
{Gavazzi} G.,  {Boselli} A.,  {Donati} A.,  {Franzetti} P.,   {Scodeggio} M.,
  2003, \mn@doi [\aap] {10.1051/0004-6361:20030026}, \href
  {http://adsabs.harvard.edu/abs/2003A\%26A...400..451G} {400, 451}

\bibitem[\protect\citeauthoryear{{Gillis} et~al.,}{{Gillis}
  et~al.}{2013}]{Gillis13}
{Gillis} B.~R.,  et~al., 2013, \mn@doi [\mnras] {10.1093/mnras/stt274}, \href
  {http://adsabs.harvard.edu/abs/2013MNRAS.431.1439G} {431, 1439}

\bibitem[\protect\citeauthoryear{{Gnedin}, {Kravtsov}, {Klypin}  \&
  {Nagai}}{{Gnedin} et~al.}{2004}]{Gnedin04}
{Gnedin} O.~Y.,  {Kravtsov} A.~V.,  {Klypin} A.~A.,   {Nagai} D.,  2004,
  \mn@doi [\apj] {10.1086/424914}, \href
  {http://adsabs.harvard.edu/abs/2004ApJ...616...16G} {616, 16}

\bibitem[\protect\citeauthoryear{{Grossauer} et~al.,}{{Grossauer}
  et~al.}{2015}]{Grossauer15}
{Grossauer} J.,  et~al., 2015, \mn@doi [\apj] {10.1088/0004-637X/807/1/88},
  \href {http://adsabs.harvard.edu/abs/2015ApJ...807...88G} {807, 88}

\bibitem[\protect\citeauthoryear{{Hahn} \& {Abel}}{{Hahn} \&
  {Abel}}{2011}]{MUSIC}
{Hahn} O.,  {Abel} T.,  2011, \mn@doi [\mnras]
  {10.1111/j.1365-2966.2011.18820.x}, \href
  {http://adsabs.harvard.edu/abs/2011MNRAS.415.2101H} {415, 2101}

\bibitem[\protect\citeauthoryear{{Haines}, {La Barbera}, {Mercurio}, {Merluzzi}
   \& {Busarello}}{{Haines} et~al.}{2006}]{Haines06}
{Haines} C.~P.,  {La Barbera} F.,  {Mercurio} A.,  {Merluzzi} P.,   {Busarello}
  G.,  2006, \mn@doi [\apjl] {10.1086/507297}, \href
  {http://adsabs.harvard.edu/abs/2006ApJ...647L..21H} {647, L21}

\bibitem[\protect\citeauthoryear{{Haines} et~al.,}{{Haines}
  et~al.}{2009}]{Haines09}
{Haines} C.~P.,  et~al., 2009, \mn@doi [\apj] {10.1088/0004-637X/704/1/126},
  \href {http://adsabs.harvard.edu/abs/2009ApJ...704..126H} {704, 126}

\bibitem[\protect\citeauthoryear{{Hayashi}, {Navarro}, {Taylor}, {Stadel}  \&
  {Quinn}}{{Hayashi} et~al.}{2003}]{Hayashi03}
{Hayashi} E.,  {Navarro} J.~F.,  {Taylor} J.~E.,  {Stadel} J.,   {Quinn} T.,
  2003, \mn@doi [\apj] {10.1086/345788}, \href
  {http://adsabs.harvard.edu/abs/2003ApJ...584..541H} {584, 541}

\bibitem[\protect\citeauthoryear{{Haynes} et~al.,}{{Haynes}
  et~al.}{2011}]{Haynes11}
{Haynes} M.~P.,  et~al., 2011, \mn@doi [\aj] {10.1088/0004-6256/142/5/170},
  \href {http://adsabs.harvard.edu/abs/2011AJ....142..170H} {142, 170}

\bibitem[\protect\citeauthoryear{{Hern{\'a}ndez-Fern{\'a}ndez}, {Haines},
  {Diaferio}, {Iglesias-P{\'a}ramo}, {Mendes de Oliveira}  \&
  {Vilchez}}{{Hern{\'a}ndez-Fern{\'a}ndez} et~al.}{2014}]{Hernandez14}
{Hern{\'a}ndez-Fern{\'a}ndez} J.~D.,  {Haines} C.~P.,  {Diaferio} A.,
  {Iglesias-P{\'a}ramo} J.,  {Mendes de Oliveira} C.,   {Vilchez} J.~M.,  2014,
  \mn@doi [\mnras] {10.1093/mnras/stt2354}, \href
  {http://adsabs.harvard.edu/abs/2014MNRAS.438.2186H} {438, 2186}

\bibitem[\protect\citeauthoryear{{Hill}, {Muzzin}, {Franx}  \&
  {Marchesini}}{{Hill} et~al.}{2017}]{Hill17}
{Hill} A.~R.,  {Muzzin} A.,  {Franx} M.,   {Marchesini} D.,  2017, \mn@doi
  [\apjl] {10.3847/2041-8213/aa951a}, \href
  {http://adsabs.harvard.edu/abs/2017ApJ...849L..26H} {849, L26}

\bibitem[\protect\citeauthoryear{{Khochfar} \& {Silk}}{{Khochfar} \&
  {Silk}}{2009}]{Khochfar09}
{Khochfar} S.,  {Silk} J.,  2009, \mn@doi [\mnras]
  {10.1111/j.1365-2966.2009.14958.x}, \href
  {http://adsabs.harvard.edu/abs/2009MNRAS.397..506K} {397, 506}

\bibitem[\protect\citeauthoryear{{Kim} \& {Park}}{{Kim} \&
  {Park}}{2006}]{Kim06}
{Kim} J.,  {Park} C.,  2006, \mn@doi [\apj] {10.1086/499761}, \href
  {http://adsabs.harvard.edu/abs/2006ApJ...639..600K} {639, 600}

\bibitem[\protect\citeauthoryear{{Klypin}, {Yepes}, {Gottl{\"o}ber}, {Prada}
  \& {He{\ss}}}{{Klypin} et~al.}{2016}]{Klypin16}
{Klypin} A.,  {Yepes} G.,  {Gottl{\"o}ber} S.,  {Prada} F.,   {He{\ss}} S.,
  2016, \mn@doi [\mnras] {10.1093/mnras/stw248}, \href
  {http://adsabs.harvard.edu/abs/2016MNRAS.457.4340K} {457, 4340}

\bibitem[\protect\citeauthoryear{{Knollmann} \& {Knebe}}{{Knollmann} \&
  {Knebe}}{2009}]{AMIGA}
{Knollmann} S.~R.,  {Knebe} A.,  2009, \mn@doi [\apjs]
  {10.1088/0067-0049/182/2/608}, \href
  {http://adsabs.harvard.edu/abs/2009ApJS..182..608K} {182, 608}

\bibitem[\protect\citeauthoryear{{Kodama}, {Balogh}, {Smail}, {Bower}  \&
  {Nakata}}{{Kodama} et~al.}{2004}]{Kodama04}
{Kodama} T.,  {Balogh} M.~L.,  {Smail} I.,  {Bower} R.~G.,   {Nakata} F.,
  2004, \mn@doi [\mnras] {10.1111/j.1365-2966.2004.08271.x}, \href
  {http://adsabs.harvard.edu/abs/2004MNRAS.354.1103K} {354, 1103}

\bibitem[\protect\citeauthoryear{{Kourkchi} \& {Tully}}{{Kourkchi} \&
  {Tully}}{2017}]{Kourkchi17}
{Kourkchi} E.,  {Tully} R.~B.,  2017, \mn@doi [\apj]
  {10.3847/1538-4357/aa76db}, \href
  {http://adsabs.harvard.edu/abs/2017ApJ...843...16K} {843, 16}

\bibitem[\protect\citeauthoryear{{Leauthaud} et~al.,}{{Leauthaud}
  et~al.}{2012}]{Leauthaud12}
{Leauthaud} A.,  et~al., 2012, \mn@doi [\apj] {10.1088/0004-637X/744/2/159},
  \href {http://adsabs.harvard.edu/abs/2012ApJ...744..159L} {744, 159}

\bibitem[\protect\citeauthoryear{Lewis, Challinor  \& Lasenby}{Lewis
  et~al.}{2000}]{CAMB}
Lewis A.,  Challinor A.,   Lasenby A.,  2000, \mn@doi [Astrophys. J.]
  {10.1086/309179}, 538, 473

\bibitem[\protect\citeauthoryear{{Li} et~al.,}{{Li} et~al.}{2014}]{Li14}
{Li} R.,  et~al., 2014, \mn@doi [\mnras] {10.1093/mnras/stt2395}, \href
  {http://adsabs.harvard.edu/abs/2014MNRAS.438.2864L} {438, 2864}

\bibitem[\protect\citeauthoryear{{Li} et~al.,}{{Li} et~al.}{2016}]{Li16}
{Li} R.,  et~al., 2016, \mn@doi [\mnras] {10.1093/mnras/stw494}, \href
  {http://adsabs.harvard.edu/abs/2016MNRAS.458.2573L} {458, 2573}

\bibitem[\protect\citeauthoryear{{Lidman} et~al.,}{{Lidman}
  et~al.}{2012}]{Lidman12}
{Lidman} C.,  et~al., 2012, \mn@doi [\mnras]
  {10.1111/j.1365-2966.2012.21984.x}, \href
  {http://adsabs.harvard.edu/abs/2012MNRAS.427..550L} {427, 550}

\bibitem[\protect\citeauthoryear{{Mahajan}, {Mamon}  \&
  {Raychaudhury}}{{Mahajan} et~al.}{2011}]{Mahajan11}
{Mahajan} S.,  {Mamon} G.~A.,   {Raychaudhury} S.,  2011, \mn@doi [\mnras]
  {10.1111/j.1365-2966.2011.19236.x}, \href
  {http://adsabs.harvard.edu/abs/2011MNRAS.416.2882M} {416, 2882}

\bibitem[\protect\citeauthoryear{{Mamon}, {Sanchis}, {Salvador-Sol{\'e}}  \&
  {Solanes}}{{Mamon} et~al.}{2004}]{Mamon04}
{Mamon} G.~A.,  {Sanchis} T.,  {Salvador-Sol{\'e}} E.,   {Solanes} J.~M.,
  2004, \mn@doi [\aap] {10.1051/0004-6361:20034155}, \href
  {http://adsabs.harvard.edu/abs/2004A%26A...414..445M} {414, 445}

\bibitem[\protect\citeauthoryear{{McDonald}, {Courteau}  \& {Tully}}{{McDonald}
  et~al.}{2009}]{McDonald09}
{McDonald} M.,  {Courteau} S.,   {Tully} R.~B.,  2009, \mn@doi [\mnras]
  {10.1111/j.1365-2966.2009.14442.x}, \href
  {http://adsabs.harvard.edu/abs/2009MNRAS.394.2022M} {394, 2022}

\bibitem[\protect\citeauthoryear{{McLaughlin}}{{McLaughlin}}{1999}]{McLaughlin99}
{McLaughlin} D.~E.,  1999, \mn@doi [\apjl] {10.1086/311860}, \href
  {http://adsabs.harvard.edu/abs/1999ApJ...512L...9M} {512, L9}

\bibitem[\protect\citeauthoryear{{Mei} et~al.,}{{Mei} et~al.}{2007}]{Mei07}
{Mei} S.,  et~al., 2007, \mn@doi [\apj] {10.1086/509598}, \href
  {http://adsabs.harvard.edu/abs/2007ApJ...655..144M} {655, 144}

\bibitem[\protect\citeauthoryear{{Merritt}, {Graham}, {Moore}, {Diemand}  \&
  {Terzi{\'c}}}{{Merritt} et~al.}{2006}]{Merritt06}
{Merritt} D.,  {Graham} A.~W.,  {Moore} B.,  {Diemand} J.,   {Terzi{\'c}} B.,
  2006, \mn@doi [\aj] {10.1086/508988}, \href
  {http://adsabs.harvard.edu/abs/2006AJ....132.2685M} {132, 2685}

\bibitem[\protect\citeauthoryear{{Navarro}, {Frenk}  \& {White}}{{Navarro}
  et~al.}{1996}]{NFW96}
{Navarro} J.~F.,  {Frenk} C.~S.,   {White} S.~D.~M.,  1996, \mn@doi [apj]
  {10.1086/177173}, \href {http://adsabs.harvard.edu/abs/1996ApJ...462..563N}
  {462, 563}

\bibitem[\protect\citeauthoryear{{Navarro}, {Frenk}  \& {White}}{{Navarro}
  et~al.}{1997}]{NFW97}
{Navarro} J.~F.,  {Frenk} C.~S.,   {White} S.~D.~M.,  1997, \mn@doi [\apj]
  {10.1086/304888}, \href {http://adsabs.harvard.edu/abs/1997ApJ...490..493N}
  {490, 493}

\bibitem[\protect\citeauthoryear{{Navarro et al.}}{{Navarro et
  al.}}{2004}]{Navarro04}
{Navarro et al.} J.~F.,  2004, \mn@doi [mnras]
  {10.1111/j.1365-2966.2004.07586.x}, \href
  {http://adsabs.harvard.edu/abs/2004MNRAS.349.1039N} {349, 1039}

\bibitem[\protect\citeauthoryear{{Niemiec} et~al.,}{{Niemiec}
  et~al.}{2017}]{Niemiec17}
{Niemiec} A.,  et~al., 2017, \mn@doi [\mnras] {10.1093/mnras/stx1667}, \href
  {http://adsabs.harvard.edu/abs/2017MNRAS.471.1153N} {471, 1153}

\bibitem[\protect\citeauthoryear{{O{\~n}orbe}, {Garrison-Kimmel}, {Maller},
  {Bullock}, {Rocha}  \& {Hahn}}{{O{\~n}orbe} et~al.}{2014}]{Onorbe14}
{O{\~n}orbe} J.,  {Garrison-Kimmel} S.,  {Maller} A.~H.,  {Bullock} J.~S.,
  {Rocha} M.,   {Hahn} O.,  2014, \mn@doi [\mnras] {10.1093/mnras/stt2020},
  \href {http://adsabs.harvard.edu/abs/2014MNRAS.437.1894O} {437, 1894}

\bibitem[\protect\citeauthoryear{{Oman}, {Hudson}  \& {Behroozi}}{{Oman}
  et~al.}{2013}]{Oman13}
{Oman} K.~A.,  {Hudson} M.~J.,   {Behroozi} P.~S.,  2013, \mn@doi [\mnras]
  {10.1093/mnras/stt328}, \href
  {http://adsabs.harvard.edu/abs/2013MNRAS.431.2307O} {431, 2307}

\bibitem[\protect\citeauthoryear{{Ouellette et al.}}{{Ouellette et
  al.}}{2017}]{Ouellette17}
{Ouellette et al.} N.~N.-Q.,  2017, \mn@doi [\apj] {10.3847/1538-4357/aa74b1},
  \href {http://adsabs.harvard.edu/abs/2017ApJ...843...74O} {843, 74}

\bibitem[\protect\citeauthoryear{{Planck Collaboration} et~al.,}{{Planck
  Collaboration} et~al.}{2018}]{Planck18}
{Planck Collaboration} et~al., 2018, preprint, \href
  {http://adsabs.harvard.edu/abs/2018arXiv180706209P} {} (\mn@eprint {arXiv}
  {1807.06209})

\bibitem[\protect\citeauthoryear{{Rhee}, {Smith}, {Choi}, {Yi}, {Jaff{\'e}},
  {Candlish}  \& {S{\'a}nchez-J{\'a}nssen}}{{Rhee} et~al.}{2017}]{Rhee17}
{Rhee} J.,  {Smith} R.,  {Choi} H.,  {Yi} S.~K.,  {Jaff{\'e}} Y.,  {Candlish}
  G.,   {S{\'a}nchez-J{\'a}nssen} R.,  2017, \mn@doi [\apj]
  {10.3847/1538-4357/aa6d6c}, \href
  {http://adsabs.harvard.edu/abs/2017ApJ...843..128R} {843, 128}

\bibitem[\protect\citeauthoryear{{Rines}, {Geller}, {Kurtz}  \&
  {Diaferio}}{{Rines} et~al.}{2005}]{Rines05}
{Rines} K.,  {Geller} M.~J.,  {Kurtz} M.~J.,   {Diaferio} A.,  2005, \mn@doi
  [\aj] {10.1086/433173}, \href
  {http://adsabs.harvard.edu/abs/2005AJ....130.1482R} {130, 1482}

\bibitem[\protect\citeauthoryear{{Roediger} \& {Courteau}}{{Roediger} \&
  {Courteau}}{2015}]{Roediger15}
{Roediger} J.~C.,  {Courteau} S.,  2015, \mn@doi [\mnras]
  {10.1093/mnras/stv1499}, \href
  {http://adsabs.harvard.edu/abs/2015MNRAS.452.3209R} {452, 3209}

\bibitem[\protect\citeauthoryear{{Roediger}, {Courteau}, {McDonald}  \&
  {MacArthur}}{{Roediger} et~al.}{2011a}]{Roediger11a}
{Roediger} J.~C.,  {Courteau} S.,  {McDonald} M.,   {MacArthur} L.~A.,  2011a,
  \mn@doi [\mnras] {10.1111/j.1365-2966.2011.19176.x}, \href
  {http://adsabs.harvard.edu/abs/2011MNRAS.416.1983R} {416, 1983}

\bibitem[\protect\citeauthoryear{{Roediger}, {Courteau}, {MacArthur}  \&
  {McDonald}}{{Roediger} et~al.}{2011b}]{Roediger11b}
{Roediger} J.~C.,  {Courteau} S.,  {MacArthur} L.~A.,   {McDonald} M.,  2011b,
  \mn@doi [\mnras] {10.1111/j.1365-2966.2011.19177.x}, \href
  {http://adsabs.harvard.edu/abs/2011MNRAS.416.1996R} {416, 1996}

\bibitem[\protect\citeauthoryear{{Ryden} \& {Gunn}}{{Ryden} \&
  {Gunn}}{1987}]{Ryden87}
{Ryden} B.~S.,  {Gunn} J.~E.,  1987, \mn@doi [\apj] {10.1086/165349}, \href
  {http://adsabs.harvard.edu/abs/1987ApJ...318...15R} {318, 15}

\bibitem[\protect\citeauthoryear{{S{\'a}nchez et al.}}{{S{\'a}nchez et
  al.}}{2012}]{CALIFA}
{S{\'a}nchez et al.} S.~F.,  2012, \mn@doi [\aap]
  {10.1051/0004-6361/201117353}, \href
  {http://adsabs.harvard.edu/abs/2012A\%26A...538A...8S} {538, A8}

\bibitem[\protect\citeauthoryear{{Sif{\'o}n} et~al.,}{{Sif{\'o}n}
  et~al.}{2015}]{Sifon15}
{Sif{\'o}n} C.,  et~al., 2015, \mn@doi [\mnras] {10.1093/mnras/stv2051}, \href
  {http://adsabs.harvard.edu/abs/2015MNRAS.454.3938S} {454, 3938}

\bibitem[\protect\citeauthoryear{{Sif{\'o}n}, {Herbonnet}, {Hoekstra}, {van der
  Burg}  \& {Viola}}{{Sif{\'o}n} et~al.}{2018}]{Sifon18}
{Sif{\'o}n} C.,  {Herbonnet} R.,  {Hoekstra} H.,  {van der Burg} R.~F.~J.,
  {Viola} M.,  2018, \mn@doi [\mnras] {10.1093/mnras/sty1161}, \href
  {http://adsabs.harvard.edu/abs/2018MNRAS.478.1244S} {478, 1244}

\bibitem[\protect\citeauthoryear{{Springel}}{{Springel}}{2005}]{GADGET2}
{Springel} V.,  2005, \mn@doi [\mnras] {10.1111/j.1365-2966.2005.09655.x},
  \href {http://adsabs.harvard.edu/abs/2005MNRAS.364.1105S} {364, 1105}

\bibitem[\protect\citeauthoryear{{Thomas}, {Saglia}, {Bender}, {Thomas},
  {Gebhardt}, {Magorrian}, {Corsini}  \& {Wegner}}{{Thomas}
  et~al.}{2009}]{Thomas09}
{Thomas} J.,  {Saglia} R.~P.,  {Bender} R.,  {Thomas} D.,  {Gebhardt} K.,
  {Magorrian} J.,  {Corsini} E.~M.,   {Wegner} G.,  2009, \mn@doi [\apj]
  {10.1088/0004-637X/691/1/770}, \href
  {http://adsabs.harvard.edu/abs/2009ApJ...691..770T} {691, 770}

\bibitem[\protect\citeauthoryear{{Tissera}, {White}, {Pedrosa}  \&
  {Scannapieco}}{{Tissera} et~al.}{2010}]{Tissera10}
{Tissera} P.~B.,  {White} S.~D.~M.,  {Pedrosa} S.,   {Scannapieco} C.,  2010,
  \mn@doi [\mnras] {10.1111/j.1365-2966.2010.16777.x}, \href
  {http://adsabs.harvard.edu/abs/2010MNRAS.406..922T} {406, 922}

\bibitem[\protect\citeauthoryear{{Tollet} et~al.,}{{Tollet}
  et~al.}{2016}]{Tollet16}
{Tollet} E.,  et~al., 2016, \mn@doi [\mnras] {10.1093/mnras/stv2856}, \href
  {http://adsabs.harvard.edu/abs/2016MNRAS.456.3542T} {456, 3542}

\bibitem[\protect\citeauthoryear{{Wechsler}, {Bullock}, {Primack}, {Kravtsov}
  \& {Dekel}}{{Wechsler} et~al.}{2002}]{Wechsler02}
{Wechsler} R.~H.,  {Bullock} J.~S.,  {Primack} J.~R.,  {Kravtsov} A.~V.,
  {Dekel} A.,  2002, \mn@doi [\apj] {10.1086/338765}, \href
  {http://adsabs.harvard.edu/abs/2002ApJ...568...52W} {568, 52}

\bibitem[\protect\citeauthoryear{{White} \& {Rees}}{{White} \&
  {Rees}}{1978}]{White78}
{White} S.~D.~M.,  {Rees} M.~J.,  1978, \mn@doi [\mnras]
  {10.1093/mnras/183.3.341}, \href
  {http://adsabs.harvard.edu/abs/1978MNRAS.183..341W} {183, 341}

\bibitem[\protect\citeauthoryear{{Zhao}, {Jing}, {Mo}  \& {B{\"o}rner}}{{Zhao}
  et~al.}{2009}]{Zhao09}
{Zhao} D.~H.,  {Jing} Y.~P.,  {Mo} H.~J.,   {B{\"o}rner} G.,  2009, \mn@doi
  [\apj] {10.1088/0004-637X/707/1/354}, \href
  {http://adsabs.harvard.edu/abs/2009ApJ...707..354Z} {707, 354}

\bibitem[\protect\citeauthoryear{{de Vaucouleurs}}{{de
  Vaucouleurs}}{1961}]{deVaucouleurs61}
{de Vaucouleurs} G.,  1961, \mn@doi [\apjs] {10.1086/190064}, \href
  {http://adsabs.harvard.edu/abs/1961ApJS....6..213D} {6, 213}

\makeatother
\end{thebibliography}


\appendix

\section{Radii and mean densities in the analytic model}
\label{appendix:radii}

\begin{figure*}
\centering
\vspace{-1.5cm}
\begin{minipage}[b]{.4\textwidth}
	\includegraphics[width=\columnwidth]{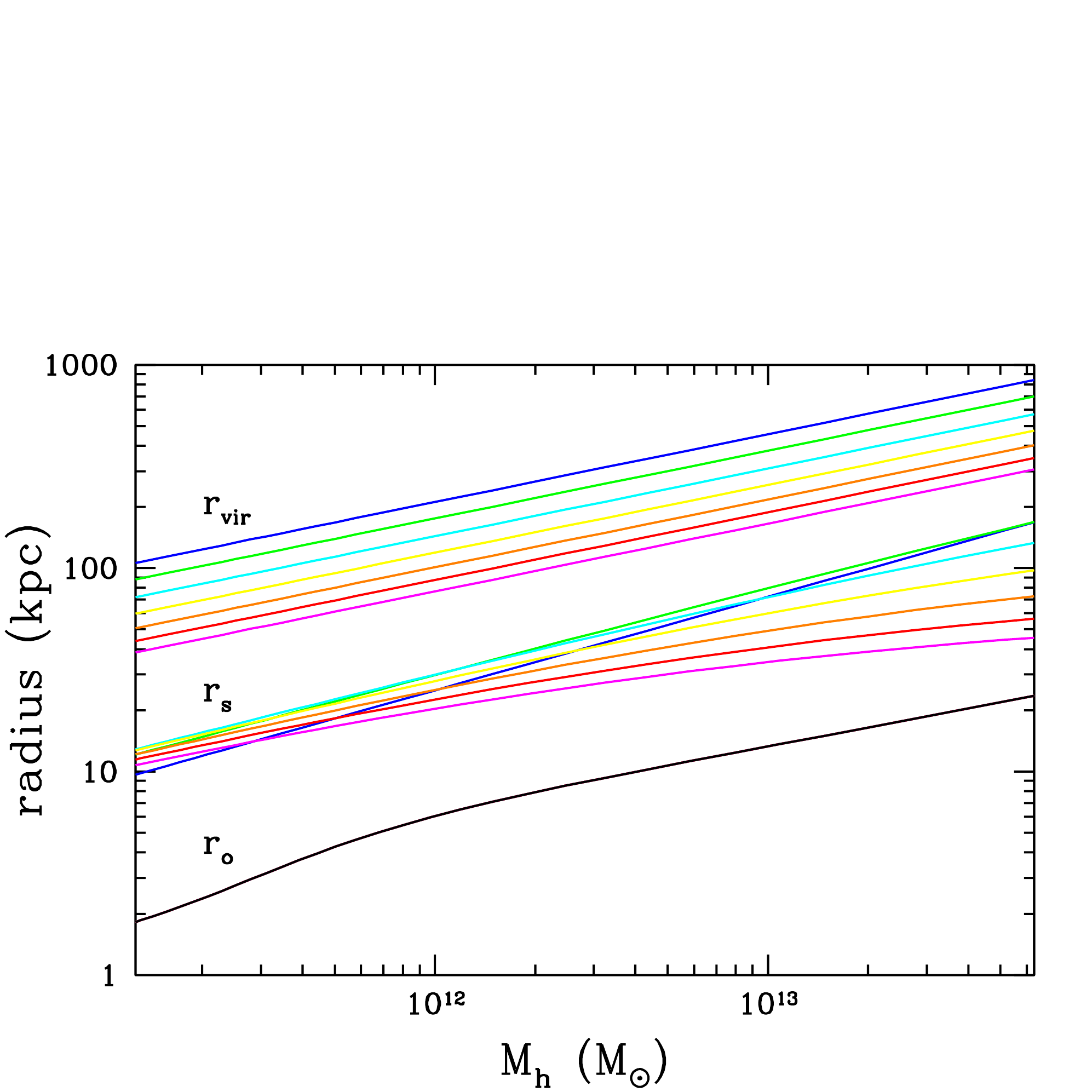}
\caption{Charactersitic radii as a function of halo mass, in the analytic model of Section \ref{subsec:2.2}. The bottom curve shows the outer radius of the galaxy $r_o$. The middle set of curves show the NFW scale radius $r_s$, for redshifts $z=0$, 0.5, 1.0, 1.5, 2.0, 2.5 and 3.0 (from top to bottom at { the} high mass end). The upper set of curves show the virial radius $r_{vir}$ for the same set of redshifts.} \label{fig:A1}
\end{minipage}\qquad
\begin{minipage}[b]{.4\textwidth}
	\includegraphics[width=\columnwidth]{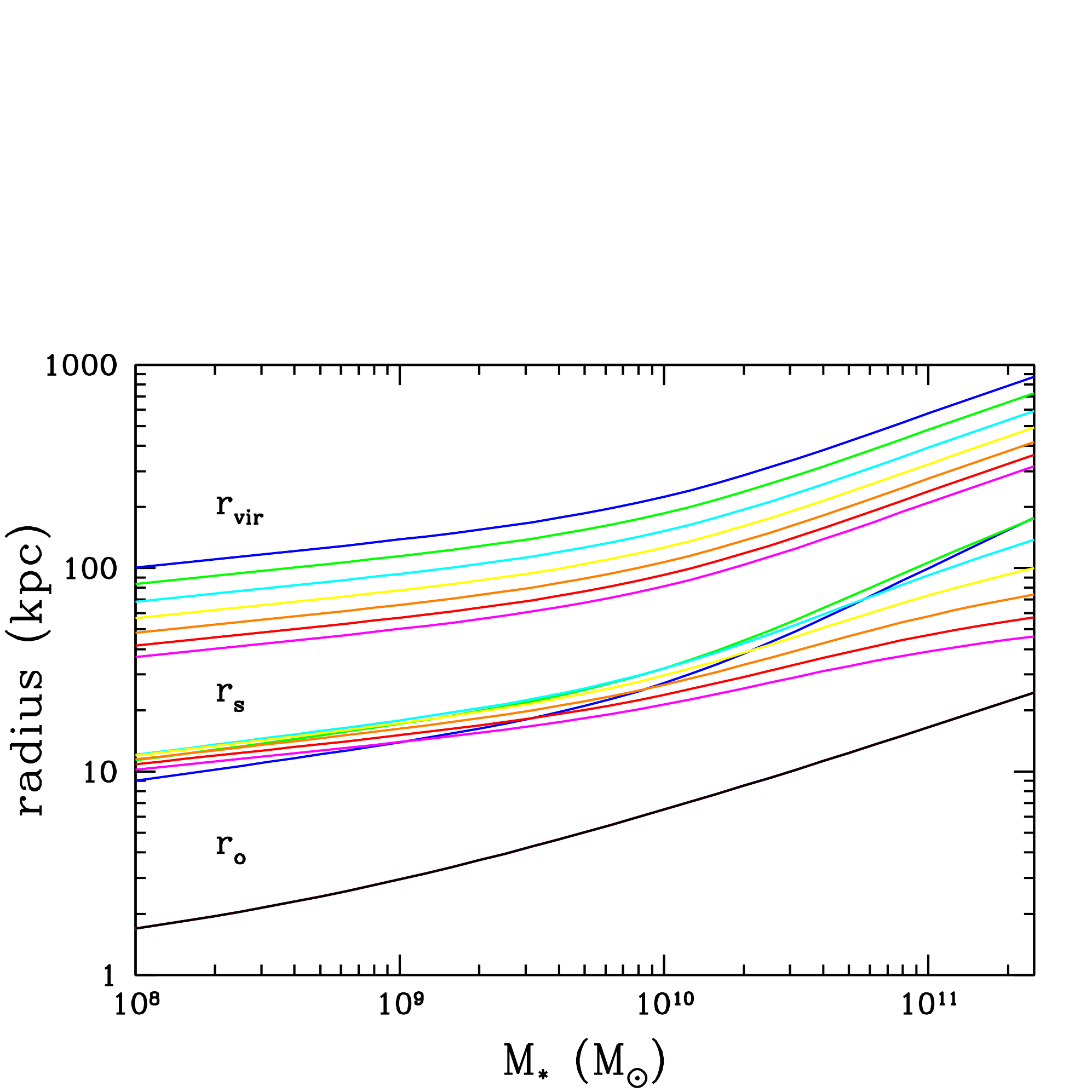}
\caption{Characteristic radii as a function of stellar mass, in the analytic model of Section \ref{subsec:2.2}. Curves are as in figure \ref{fig:A1}.}\label{fig:A2}
\vspace{0.67truein}
\end{minipage}
\end{figure*}

For clarity and to aid physical intuition, we plot some of the characteristic scales and densities in the analytic model of Section \ref{subsec:2.2}, as a function of halo mass and stellar mass. Figs.~\ref{fig:A1} and \ref{fig:A2} show the outer radius $r_{\rm o}$, scale radius $r_{\rm s}$ and virial radius $r_{\rm vir}$ as a function of halo mass and of stellar mass respectively, for redshifts  $z=0$, 0.5, 1.0, 1.5, 2.0, 2.5 and 3.0. While the virial radius simply scales as $M_{\rm h}^{1/3}$, the scale radius has a more complicated behaviour due to the varying concentration-mass-redshift relation. The outer radius $r_o$ is always smaller than the scale radius, and generally lies in the range $r_{\rm o}$ 0.1--0.2 $r_{\rm s}$. The kink in the relationship between halo mass and stellar mass further distorts these patterns in Fig.~\ref{fig:A2}.

\begin{figure*}
\centering
\vspace{-2.5cm}
\begin{minipage}[b]{.4\textwidth}
	\includegraphics[width=\columnwidth]{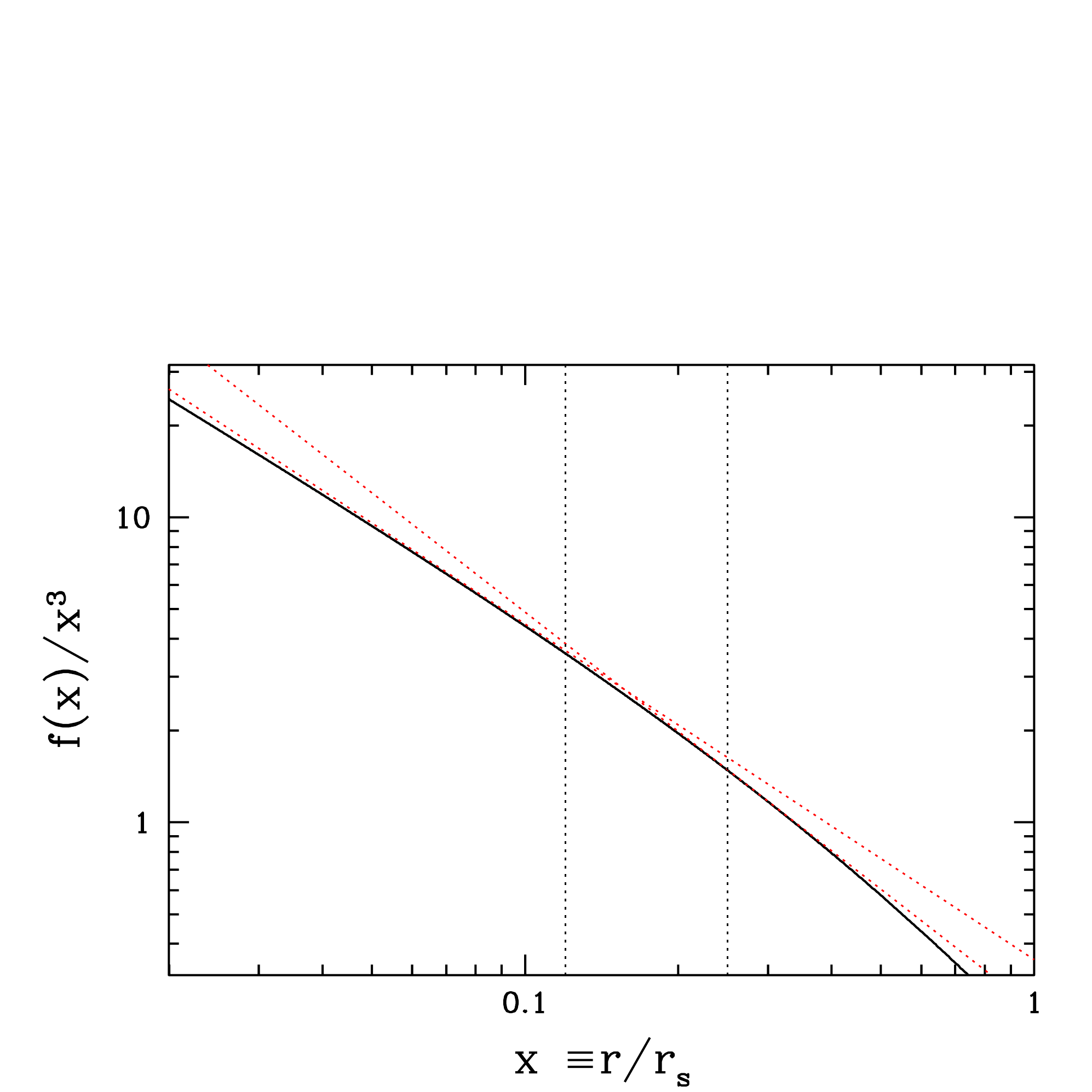}
\caption{NFW mean density $f(x)/x^3$ as a function of $x \equiv r/r_s$. Dotted vertical lines indicate the upper and lower range of typical values for $x_{\rm o} = r_{\rm o}/r_{\rm s}$. (Dotted red) lines of slope $-1.1$ and $-1.3$ are also drawn.}\label{fig:A3}
\end{minipage}\qquad
\begin{minipage}[b]{.4\textwidth}
	\includegraphics[width=\columnwidth]{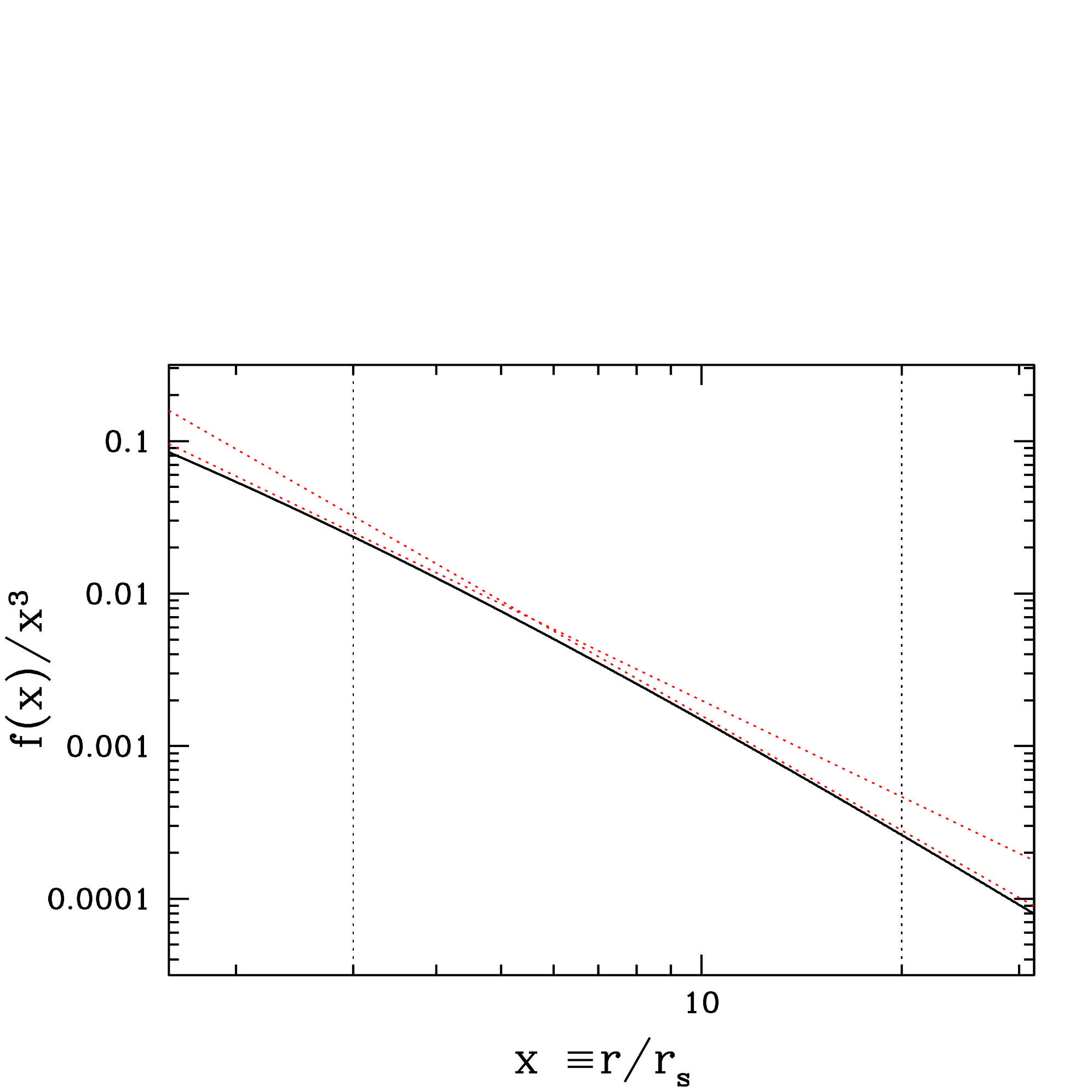}
\caption{As in figure \ref{fig:A3}, but for the range typical of the virial radius. The vertical lines show the upper and lower range of typical values of $c = r_{\rm vir}/r_{\rm s}$,
while dotted (red) lines of slope $-2.1$ and $-2.5$ are also shown.}\label{fig:A4}
\end{minipage}
\vspace{-0.2cm}
\end{figure*}

Figs \ref{fig:A3} and \ref{fig:A3} show how { the} mean density function $f(x)/x^3$ scales with radius, around the values $x \equiv r/r_{\rm s} = 0.1$--0.2 and $x \equiv r/r_{\rm s} = 3$--20 respectively. These are the ranges at which the outer radius $r_{\rm o}$ and the virial radius $r_{\rm vir}$ typically lie. Lines indicate slopes of $-1.1$ to $-1.3$, and $-2.1$ to $-2.5$ on the two figures respectively.

\begin{figure*}
\centering
\begin{minipage}[b]{.4\textwidth}
	\includegraphics[width=\columnwidth]{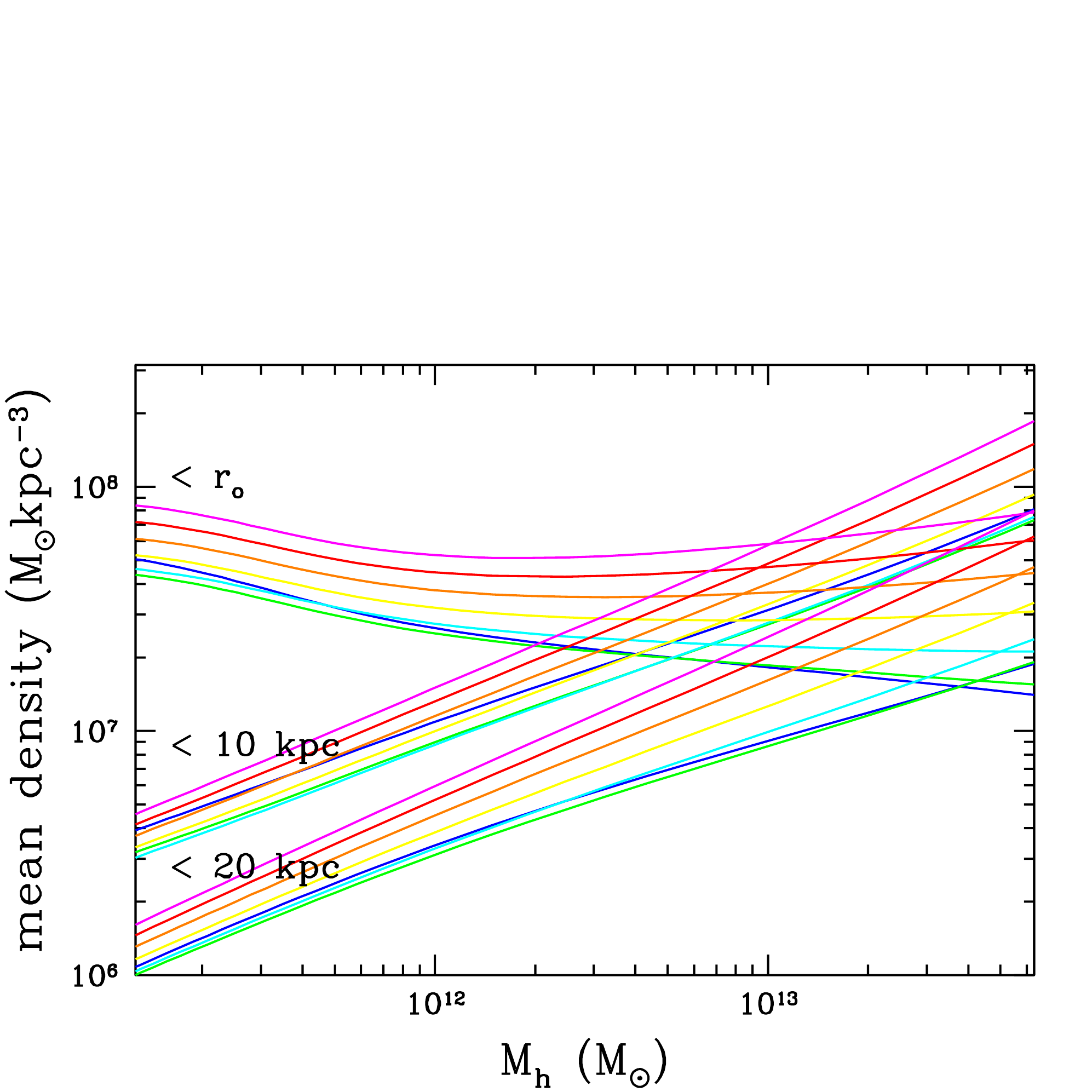}
\caption{Mean densities as a function of halo mass, in the analytic model of Section \ref{subsec:2.2}. The bottom set of curves show the mean density within a fixed radius of 20 kpc,  for redshifts $z=0$, 0.5, 1.0, 1.5, 2.0, 2.5 and 3.0 (from bottom to top at high mass end). The middle set of curves show the mean density within a fixed radius of 10 kpc for the same set of redshifts, while the top set of curves show the mean density within $r_{\rm o}$.}\label{fig:A5}
\end{minipage}\qquad
\begin{minipage}[b]{.4\textwidth}
	\includegraphics[width=\columnwidth]{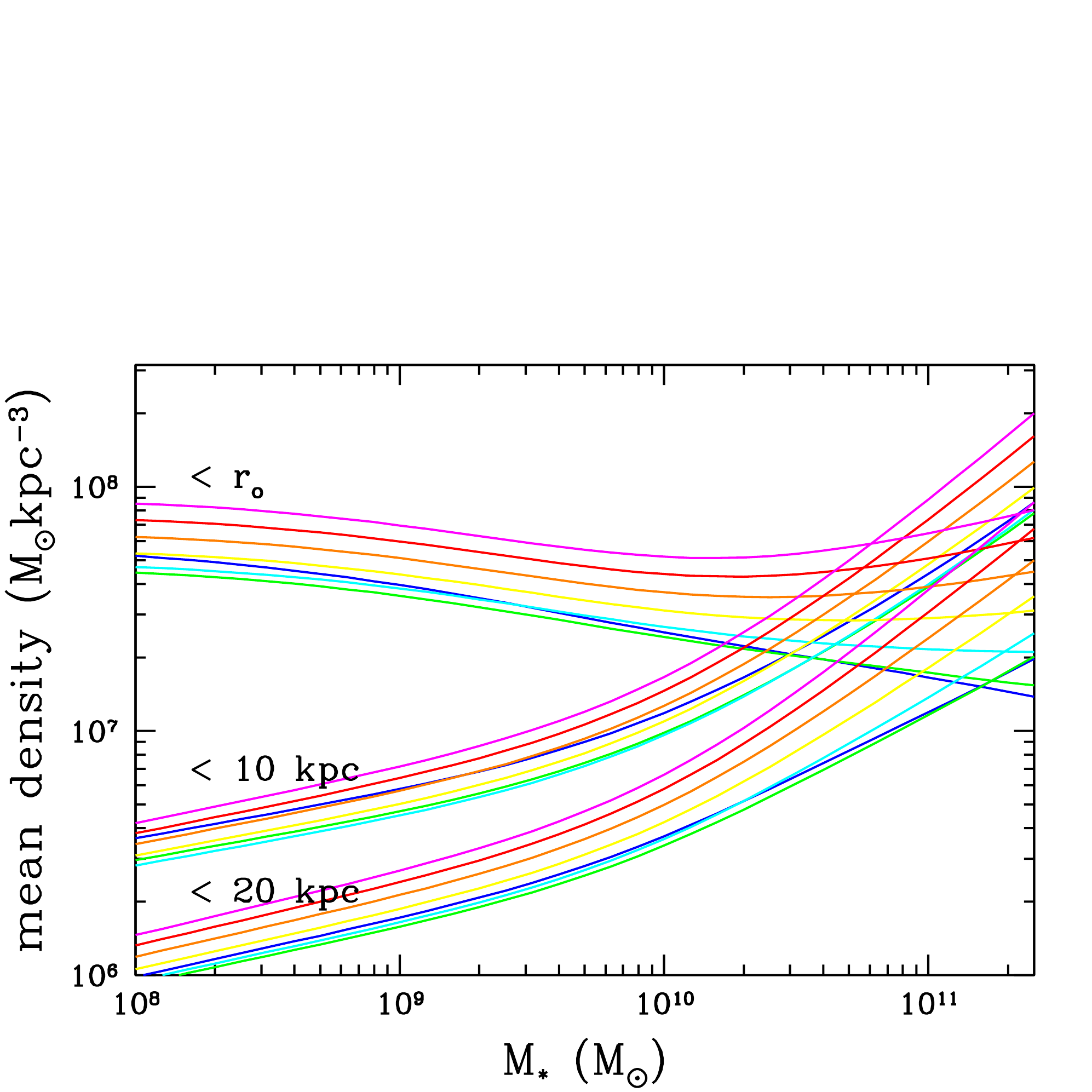}
\caption{Mean densities as a function of stellar mass, in the analytic model of Section \ref{subsec:2.2}. Curves are as in Fig.~\ref{fig:A5} }\label{fig:A6}
\vspace{0.7truein}
\end{minipage}
\end{figure*}

Figs.~\ref{fig:A5} and \ref{fig:A6} show the mean density enclosed within fixed radii of 10 kpc or 20 kpc, as well as the density within the outer radius radius $r_{\rm o}$, as a function of halo mass and of stellar mass respectively, for redshifts  $z=0$, 0.5, 1.0, 1.5, 2.0, 2.5 and 3.0. While the mean density within a fixed physical radius increases with halo mass or stellar mass, it is always lower at the larger radius. Since the outer radius $r_{\rm o}$ increases with halo mass or stellar mass, the two effects cancel almost completely, and the mean density within $r_{\rm o}$ is almost independent of halo mass or stellar mass, declining by a factor of only 3 or less, over a range of more than 500 in halo mass, or more than 2000 in stellar mass.

\section{Net Effect of Adiabatic Contraction}
\label{appendix:contraction}

 In the classic calculation of adiabatic contraction \cite{Blumenthal86}, it is assumed that particle orbits are circular and that shells of matter do not cross as the system contracts. The adiabatic invariant for a given shell of radius $r$, enclosing mass $M(r)$, is then $rM(r)$, 
To solve for the final dark matter distribution, we can write the conservation equation as
\begin{eqnarray}
 r_f M_{{\rm tot},f}( r_f) - r_i M_{{\rm tot},i}(r_i) = 0&&\\
 \Leftrightarrow r_f\left[M_{b,f}(r_f) + M_{d,f}(r_f)\right]  - r_i\left[M_{{\rm tot},i}(r_i)\right] = 0&&
\end{eqnarray}
where $r_i$ and $r_f$ are the initial and final radius of the shell, $M_{{\rm tot},f}$ and $M_{{\rm tot},i}$ are the initial and final total mass within the shell, and $M_b$ and $M_d$ are the baryonic and dark matter components respectively. If we assume that the baryons trace the dark matter initially, then we can write $M_{{\rm tot},i}(r_i) = M_{d,i}(r_i)/(1-f_b)$, where $f_b$ is the baryon fraction in the initial conditions. Furthermore, since we have assumed no shell crossing, $M_{d,i}(r_i) = M_{d,f}(r_f)$, and thus
\begin{eqnarray}
 r_f M_{b,f}(r_f) + r_f M_{d,f}(r_f) - r_i M_{d,i}(r_i)/(1-f_b) =&& \\
 r_f M_{b,f}(r_f) + r_f M_{d,i}(r_i) - r_i M_{d,i}(r_i)/(1-f_b)  = 0&&
\end{eqnarray}

Given a form of the initial matter distribution (e.g. an unmodified NFW profile) and the final baryon distribution (e.g. a Sersic profile), we can solve this implicit equation for $r_i(r_f)$, and obtain the contraction factor $r_f/r_i$. In general, if we consider the radius in the final system at which the stellar-to-dark-matter ratio is $R_{bd}$, then 
\begin{eqnarray}
r_f R_{bd} M_{d,f}(r_f) + r_f M_{d,i}(r_i) - r_i M_{d,i}(r_i)/(1-f_b)= \\
r_f R_{bd} M_{d,i}(r_i) + r_f M_{d,i}(r_i) - r_i M_{d,i}(r_i)/(1-f_b)= 0\\
\Leftrightarrow r_fR_{bd} + r_f - r_i/(1-f_b) =0\\
\Leftrightarrow r_i/r_f = (R_{bd} + 1)(1-f_b)
\end{eqnarray}
while the \cite{Gnedin04} model predicts 
\begin{equation}
r_i/r_f = [(1+R_{bd})(1+f_b)]^{0.8}.
\end{equation}

\section{Density versus Position in Simulated Subhalos}
\label{appendix:simgradient}

\begin{figure}
	\includegraphics[width=\columnwidth]{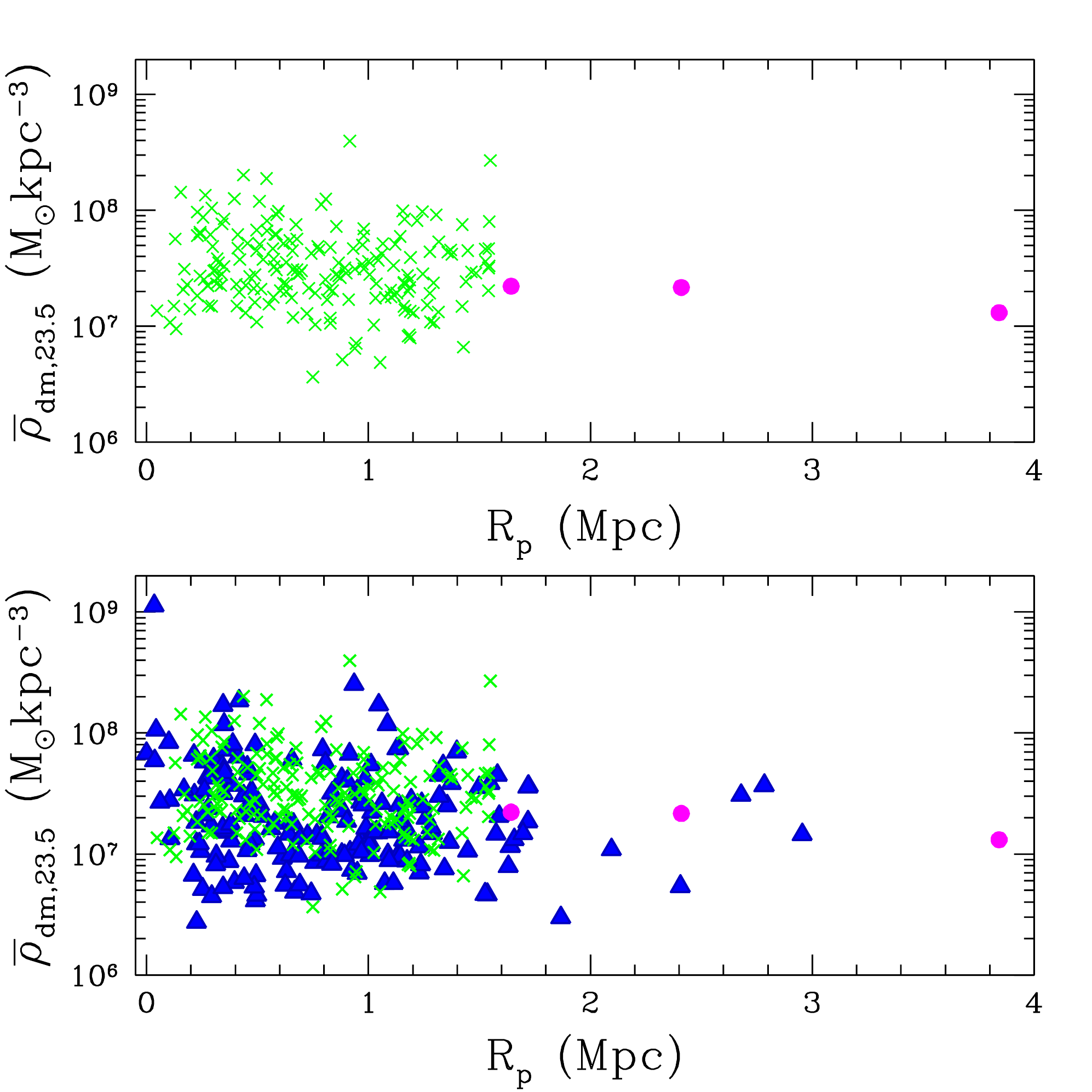}
    \vspace{-0cm}
    \caption{{{\it Top panel:} Mean dark matter density within $r_{\rm o}$ versus projected position within the cluster, as in Fig.~\ref{fig:7}, for subhalos in one of the simulated clusters (green crosses), as well as for a few nearby field halos in the high-resolution volume (magenta circles). {\it Bottom panel:} The simulation results, compared to the distribution of density versus $R_{p,M87}$ in the SHIVRir sample (blue triangles).}}
    \label{fig:A7}
\end{figure}

{ The top panel of Fig.~\ref{fig:A7} shows the mean dark matter density within $r_{\rm o}$ versus projected position within the cluster, as in Fig.~\ref{fig:7}, for subhalos in one of the simulated clusters (green crosses; note that the subhalos are all within the virial radius by definition, given our group finding algorithm.) The magenta circles show densities in a few field halos that lie close to the cluster within the high-resolution region of the simulation. The bottom panel shows the same points, compared with the distribution of density versus $R_{p,M87}$ in the SHIVRir sample (blue triangles). Overall, there is reasonable agreement between the two distributions. While there is a slight radial gradient in dark matter density in the simulation results, it is less apparent than in Virgo, possibly because the simulated cluster has experienced recent major mergers, as discussed in Section \ref{subsec:2.3}, and therefore has a less clearly stratified structure.}


\bsp	
\label{lastpage}
\end{document}